\documentclass[usenatbib,hidelinks]{mnras}
\usepackage{amsmath, placeins,xcolor,
hyperref,graphicx,pbox,epstopdf}

\graphicspath{{./figures/}}

\newcommand{\nhat}{\hat{\mathbf{n}}}

\setcitestyle{citesep={,}}

\title[Detectability of Galactic Faraday Rotation]{Detectability of Galactic Faraday Rotation in multiwavelength CMB observations \\ 
 {\footnotesize A cross-correlation analysis of CMB and radio maps}}

\author[Kolopanis, Mauskopf, \& Bowman]{
Matthew Kolopanis,${}^{1,2}$
Philip Mauskopf,${}^{1,2}$
Judd Bowman${}^{2}$\\
${}^{1}$Department of Physics, Arizona State University, Tempe, AZ 85287\\
${}^{2}$School of Earth and Space Exploration, Arizona State University, Tempe, AZ 85287 
}

\begin{document}
\maketitle
\begin{abstract} 
We introduce a new cross-correlation method to detect and verify the astrophysical origin of Faraday Rotation (FR) in multiwavelength surveys.
FR is well studied in radio astronomy from radio point sources 
but the $\lambda^{2}$ suppression of FR
makes detecting and accounting for this effect difficult at millimeter and sub-millimeter wavelengths. 
Therefore statistical methods are used to attempt to detect FR in the cosmic microwave background (CMB). Most estimators of the FR power spectrum rely on single frequency data. 
In contrast, we investigate the correlation of polarized CMB maps with FR measure maps from radio point sources.
We show a factor of $\sim30$ increase in sensitivity over single frequency estimators and predict detections exceeding $10\sigma$ significance for a CMB-S4 like experiment.
Improvements in observations of FR from current and future radio polarization surveys will greatly increase the usefulness of this method.
 
\end{abstract}

\section{Introduction}\label{sec:intro}{
 Current and future polarized cosmic microwave background (CMB) experiments like 
PLANCK \citep{planck:2015i}, 
QUIET \citep{quiet_instrument}, 
WMAP \citep{wmap:2013}, 
CLASS \citep{class:2014}, 
SPT \citep{spt3g:2014}, 
SPIDER \citep{spider:2013}, 
and The BICEP/KECK array \citep{bicep_keck:2014,bicepbmode:2015} image the cosmic microwave background with increasing sensitivity.
 In particular, these experiments are improving the sensitivity to the polarized E-mode signal and providing better wavelength coverage (30--220~GHz) compared to previous generations of CMB experiments.
  The first B-mode signals have also been detected via 
 the lensing B-modes in a cross-correlation from 
 SPT \citep{sptbmode:2013} and ACT \citep{actbmode:2015}, 
 autocorrelation from SPTpol \citep{keisler:2015}, 
 ACTpol \citep{Naess:2014} and Polarbear \citep{polarbear:2014}, 
 and dust generated B-modes in auto and cross-correlation \citep{bicepbmode:2015,planck_bicep:2015}. 

  Another foreground contribution to the B-mode signal in the CMB is Faraday Rotation (hereby referred to as FR) \citep{scoccola:2004,tashiro:2008}. Primordial, Galactic and extragalactic contributions to FR will cause E-mode and B-mode mixing \citep{Gluscevic:2009}. Future B-mode experiments will need to remove this signal in order to accurately characterize polarized signals from primordial sources or an EB cross-correlation.  
 
 FR is the displacement of the polarization angle of linearly polarized photons as they propagate through a plasma.
  While dust or synchrotron polarization provides information on the component of magnetic fields oriented perpendicular to the line of sight, FR is a probe of magnetic fields along the line of sight.
  
 The presence of `isotropic birefringence' can also rotate the polarization angle of linearly polarized photons. 
This effect, however, manifests in a frequency independent manner and imprints a unique, $\ell$-independent, signature on the CMB power spectrum. 
In contrast, the anisotropic nature of FR imprints a signature with known $\ell$-dependent structure on CMB power spectra.  \citep{lue:1999,gruppuso:2016}

  Available maps of Galactic and extragalactic FR measure, shown in \autoref{fig: Radio RM}, rely on current radio data \citep{Oppermann:2014} but upcoming radio surveys will provide better overall sensitivity and more precise measurements of FR \citep{bernardi:2013,sotomayor-beltran:2013,condon:2015,sotomayor-beltran:2015, wayth:2015, lenc:2016}.
  
 FR becomes significant for photons travelling through regions with large magnetic fields oriented parallel to the direction of photon propagation and regions with weak magnetic fields extending over non-trivial distances \citep{De:2013}.
  \begin{figure}[H!t]
 \centering
 \includegraphics[trim= 30ex 1.5ex 30ex 1ex,clip=false, scale=.40]{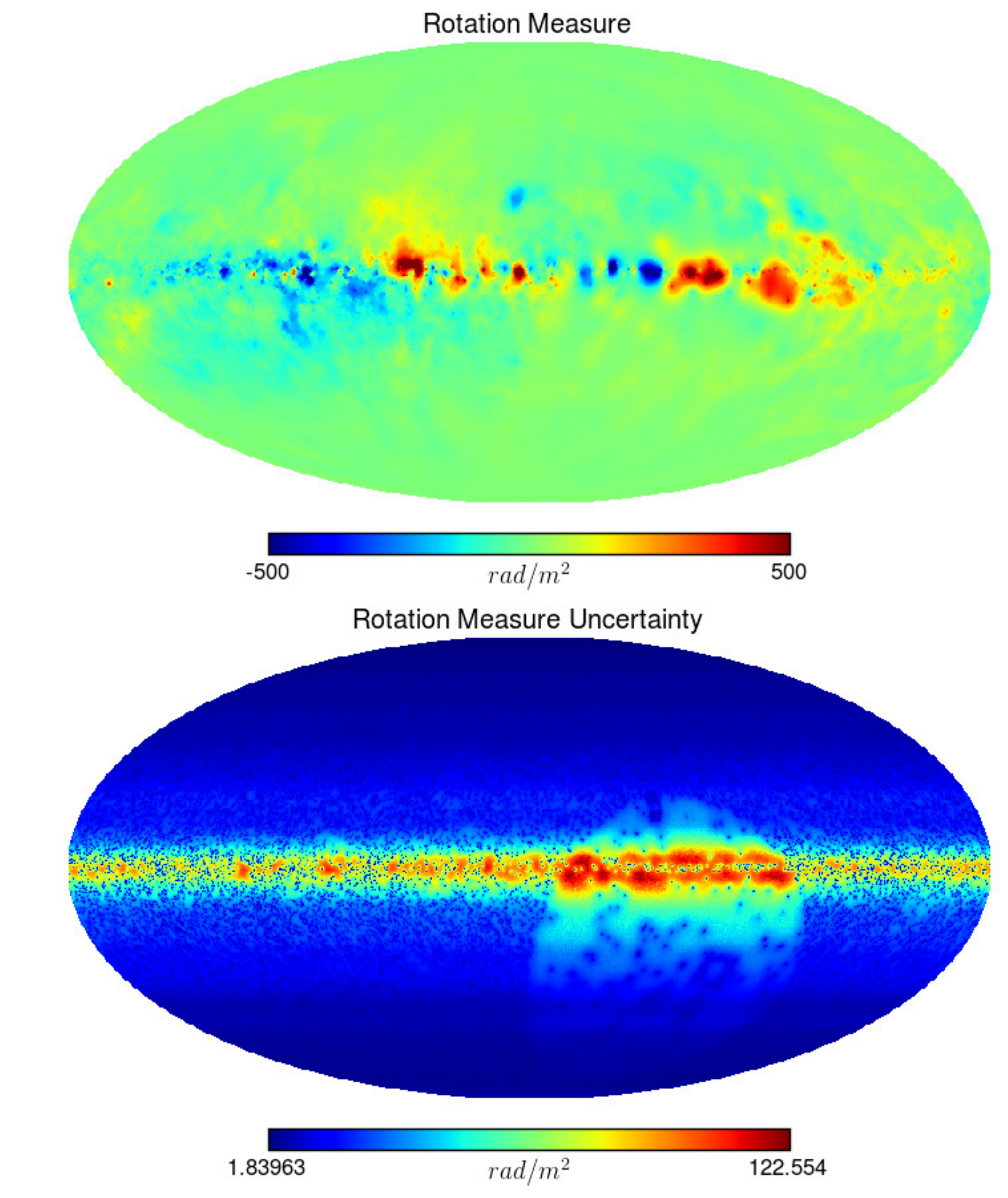}
 \caption{RM of the Galaxy as provided by \protect\cite{Oppermann:2014}. The top figure is the RM reconstruction and the bottom is uncertainty in RM. Note the difference in scales. \label{fig: Radio RM}	}
 \end{figure}
 
 In near-large galaxies, the high electron density, $n_{e}$, can cause large FR and may also contain tangled magnetic fields that can lead to depolarization \citep{spass:2010}. 
  Depolarization is the net loss of the 
 total polarized intensity. FR can also cause depolarization
 through differential FR. Differential FR occurs when polarized photons are emitted from a spatially extended source or, in the case of the CMB, from a large primordial magnetic field at the surface of last scattering. Photons undergo different amounts of FR depending upon the extent of the source through which they travel. 
 
For the CMB specific case, a large magnetic field at the surface of last scattering will create a damping effect on E- and B-mode production.
 In the presence of a large magnetic field, Thomson scattering at the surface of last scattering will cause depolarization \citep{scoccola:2004}.  Specific discussion of these effects can be found in \citet{harari:1997} and \citet{scoccola:2004}  respectively. 

   The detection of FR at millimeter wavelengths would enable
the detection of the rotation of polarized CMB emission due to interactions at high redshifts (e.g. reionization or recombination epochs). 
This would potentially constrain the amplitude of large-scale magnetic fields.

Characterizing FR through different cosmological eras  will also provide insight to the evolution of magnetic fields in the Universe.
  Observable FR first occurred during photon decoupling at the surface of last scattering \citep{Kosowsky:1996yc}. 
  When the universe is still at a high ionized fraction, photons that have decoupled from the baryionic fluid will experience FR while recombination occurs. Reionization will also leave a signature of FR as ionization fractions increase and photons pass through ionized regions \citep{scoccola:2004}. 
   
  Predictions for FR in the CMB at recombination from primordial magnetic fields estimated a $1^{\circ}$ rotation in polarization angle at an observed frequency of 30~GHz \citep{Kosowsky:1996yc}. The resulting power spectrum of these polarized photons is estimated to have a peak polarization amplitude of 
$\ell^{2}C_{\ell}^{C} \approx 10^{-12} (\mu K)^{2}$ \citep{kosowsky:2005}. 
 Recent estimates of the strength of primordial magnetic fields from the PLANCK collaboration correspond to a level of FR in the CMB comparable to the amount expected from Galactic sources \citep{planck:2015pmf}. A precise understanding of Galactic FR is required to disentangle the two signals. Detecting Galactic FR in CMB data sets is the first step in this process.	

 The paper is organized as follows: Section~\ref{sec:faraday rotation} reviews FR, 
  Section~\ref{sec:correlation detection} defines the correlator used in this paper and its uncertainty, 
  Section~\ref{sec:simulation} analyses the simulations of this correlator,  
  Section~\ref{sec:foregrounds} explores the correlation of FR with other CMB foregrounds, 
 Section~\ref{sec:real-data} applies this analysis to real data, 
and we provide a discussion of this work in  Section~\ref{sec:summary}.
}

\section{Faraday Rotation}\label{sec:faraday rotation}{

 FR occurs as polarized photons propagate through regions of space  containing ionized particles and magnetic fields. These photons undergo a rotation in the direction of polarization by an angle 
\begin{equation}
\theta(\nhat)=\lambda^{2} \alpha_{RM}(\nhat)=\frac{3}{16\pi^{2}e}\lambda^{2} \int \dot{\tau}\mathbf{B}\cdot d\mathbf{l}  \label{eqn:fr}
\end{equation}
In this equation $\nhat$ is the direction along the line of sight, $\dot{\tau}\equiv n_{e}\sigma_{T}a$ is the differential optical depth,  $\lambda$ is the observed wavelength of the photon, $\mathbf{B}$ is the comoving magnetic  field 
and $d\mathbf{l}$ is the comoving element of length along the trajectory of the photon \citep{De:2013}. The differential optical depth is a function of the free electron density along the line of sight, $n_{e}$, the Thomson scattering cross-section, $\sigma_{T}$ , and the scalefactor, $a$.
	The rotation measure, $\alpha_{RM}(\nhat)$, is the wavelength-independent quantity describing the strength of FR along the line of sight. 
	 
Under FR, the Stokes parameters are transformed as
\begin{equation}
Q_{\lambda} + i U_{\lambda} = (Q_{0} + iU_{0})e^{2i\theta(\lambda,\nhat)}  \label{stokes_transform}
\end{equation}
where $Q_{0}$ and $U_{0}$ are the un-rotated $Q$ and $U$ parameters of the photons in the limit lambda goes to 0, 
equivalent to the intrinsic polarization of the radiator 
(e.g. the surface of last scattering).

While the un-rotated polarization bases cannot be directly observed, the effects of FR in multifrequency experiments can be observed through the phase difference between frequencies
\begin{equation}
Q_{i} + i U_{i} = (Q_{j} + iU_{j})e^{2i(\lambda_{i}^{2}-\lambda_{j}^{2})\alpha(\nhat)} \label{eqn:stokes_multi-freq}
\end{equation}
where subscripts $i,j$ represent different observation frequency bands. In other words, the polarization vector, $Q +iU$, should differ by a phase proportional to the difference of the squares of wavelengths between two frequency bands.

Estimators of the RM power spectrum can be constructed from direct observation of the Gradient (E-mode) and Curl (B-mode) power spectra of the CMB \citep{ Gluscevic:2009, Kamionkowski2008fp, Yadav2009, De:2013, pogosian:2013}.
Such an estimator can also be used to constrain the strength of primoridial magnetic fields \citep{pogosian:2013,Ade:2015cao,planck:2015pmf}. 
These optimal estimators will help constrain early universe models and the evolution of magnetic fields.
Unfortunately, noise levels in current CMB experiments are too high to characterize the FR power spectrum. 

In this paper, we calculate the cross-correlation of FR measure maps provided by \citet{Oppermann:2014} with CMB maps. 
 This correlation can be used to verify the presence of FR in the CMB from a known source (e.g. FR measured from radio observations) and as a tool to verify astrophysical FR in CMB observations. 
Since FR is cumulative, direct fitting for $\alpha_{RM}$ will only recover the net effect of astrophysical FR, 
FR intrinsic to sources
and any systematic effects that manifest in the uncertainty of polarization angle.

}

\section{Faraday Rotation Correlator}\label{sec:correlation detection}{
\subsection{Correlator}{
To construct this correlation, consider maps of observed CMB Stokes parameters $Q$ and $U$. For each pixel, $n$, in these maps, the small angle approximation of equation~\autoref{eqn:stokes_multi-freq} becomes
\begin{equation}
\begin{aligned}
Q_{i}^{n} + i U_{i}^{n} &= (Q_{j}^{n} + iU_{j}^{n})e^{2i(\lambda_{i}^{2}-\lambda_{j}^{2})\alpha(\nhat)} \label{eqn:QUsep}\\
&\approx  (Q_{j}^{n} + iU_{j}^{n})(1+2i(\lambda_{i}^{2}-\lambda_{j}^{2})\alpha_{RM}^{n}) 
\end{aligned}
\end{equation} 
 Since Stokes $Q$ and $U$ are both real quantities, we can separate the real and imaginary parts of this equation and calculate the 
difference in each as
 \begin{equation}
 \begin{aligned}
  \Delta Q_{ij}^{n}= Q_{i}^{n}-Q_{j}^{n}= 2(\lambda_{i}^{2}-\lambda_{j}^{2})\alpha_{RM}^{n} U_{i}^{n}\\
 \Delta U_{ij}^{n}= U_{i}^{n}-U_{j}^{n}= -2(\lambda_{i}^{2}-\lambda_{j}^{2})\alpha_{RM}^{n} Q_{i}^{n} \label{eqn:QUmaps}
 \end{aligned}
 \end{equation}
 where the subscripts $i,j$ represent frequency bands.
  The minus sign convention here is chosen such that the resulting power spectra are positive. Applying this correlator requires maps of $Q$ and $U$ from at least two frequencies and a sufficient map of $\alpha_{RM}$. 
  To combine more than two pairs of maps we employ an inverse 
  variance weighting of the correlation for multiple frequency pairs.
 

 Using the standard spherical harmonic decomposition, we define
\begin{equation}
   \frac{\Delta Q_{ij}}{2(\lambda_{i}^{2}-\lambda_{j}^{2})} \equiv\sum_{\ell m} q_{\ell m}^{ij}Y_{\ell m}
  \label{eqn:dqalphaalm}
\end{equation}
\begin{equation}
  \frac{\Delta U_{ij}}{2(\lambda_{i}^{2}-\lambda_{j}^{2})} \equiv\sum_{\ell m} u_{\ell m}^{ij}Y_{\ell m}
   \label{eqn:dualphaalm}
\end{equation}
\begin{equation}
  -\alpha_{RM}(\nhat) Q_{i} \equiv \sum_{\ell m} r_{\ell m}^{i}Y_{\ell m} \label{eqn:qalm}
\end{equation}
\begin{equation}
    \alpha_{RM}(\nhat)U_{i} \equiv \sum_{\ell m} s_{\ell m}^{i}Y_{\ell m}
     \label{eqn:ualm}
 \end{equation}
The factor $2\left( \lambda_{i}^{2} - \lambda_{j}^{2} \right)$ in Equations~\ref{eqn:dqalphaalm}~and~\ref{eqn:dualphaalm} is introduced to construct a wavelength-independent correlator. These four maps will form the basis of our cross-correlations. Equation~\ref{eqn:QUmaps} shows that under small rotations the quantities in equations~\ref{eqn:dqalphaalm}~and~\ref{eqn:ualm} should be equivalent
since the $2\left( \lambda_{i}^{2} - \lambda_{j}^{2} \right)$ will cancel in the definition of $\Delta Q_{ij}$ in equation~\ref{eqn:QUmaps}. This will also hold for equations~\ref{eqn:dualphaalm}~and~\ref{eqn:qalm}.

From the spherical harmonic coefficients, cross-correlations can be defined as
\begin{equation}
 C_{\ell}^{AB} =\frac{1}{2\ell +1}\sum\limits_{m=-\ell}^{\ell}Re \left\lbrace A_{\ell m}^{*}B_{\ell m} \right\rbrace  \label{eqn:correlation}
 \end{equation}
Where A and B denote the two maps used in a cross-correlation, for this work equaitons~\ref{eqn:dqalphaalm}~and~\ref{eqn:ualm}, and equations~\ref{eqn:dualphaalm}~and~\ref{eqn:qalm}.
These two angular power spectra can then be added together to create the detection correlator
\begin{equation}
\begin{aligned}
C_{\ell}^{FR}&= \frac{1}{2\ell + 1} \sum_{m=-\ell}^{\ell} Re \left\lbrace \  q_{\ell m}^{*}s_{\ell m} + \ u_{\ell m}^{*}r_{\ell m} \   \right\rbrace \\
 &= C_{\ell}^{\Delta Q \times \alpha U} + C_{\ell}^{\Delta U \times \alpha Q}  \label{eqn:correlator} 
 \end{aligned}
\end{equation}

}

\subsection{Uncertainty}\label{sec:uncertainty}{
The theoretical uncertainty in equation~\ref{eqn:correlator} can be calculated following the work of \cite{polenta:2005}. The method calculates the uncertainty in a cross-correlation as
\begin{equation}
{\delta \tilde{C}^{ij}_{\ell}}^{2} = \frac{2}{\nu_{\ell}} \left\lbrace {C_{\ell,th}^{ij}}^{2} + \frac{C^{ij}_{\ell,th}}{2} \left(N_{\ell}^{i} + N_{\ell}^{j} \right) + \frac{N_{\ell}^{i}N_{\ell}^{j}}{2} \right\rbrace  \label{eqn:cross-correlation-variance}
 \end{equation}
 where $\nu_{\ell} = (2\ell + 1)\Delta\ell f_{sky} \frac{w_{2}^{2}}{w_{4}}F_{\ell}$, $f_{sky}$ is the fraction of the sky observed, $\Delta \ell$ is the size of a bin in $\ell$ space, $w_{2}$ and $w_{4}$ are powers of integrals of a pixel space masking function and $F_{\ell}$ is a power-transfer function. These quantities and an in-depth analysis can be found in \citet{Hivon2001}. The subscript $th$ denotes a theoretical model. For this work, we represent $C^{i,j}_{\ell,th}$ as multiple realizations over noiseless CMB simulations inserted into the correlator pipeline. The two maps are given superscripts  $i \text{ and } j$, and their respective noise power spectra denoted as $N_{\ell}^{i,j}$.
 
An estimate of the noise power in equation~\ref{eqn:dqalphaalm}, $N_{\ell}^{\Delta Q}$, and equation~\ref{eqn:ualm}, $N_{\ell}^{\alpha U}$, is required to use equation~\ref{eqn:cross-correlation-variance}.
To accomplish these estimates, considering equations~\ref{eqn:dqalphaalm}~and~\ref{eqn:ualm}, write
\begin{equation} 
q^{ij}_{\ell m} = \int d\Omega \left[  \frac{Q^{i}(\theta,\phi) - Q^{j}(\theta,\phi)}{2 \left(\lambda_{i}^{2}-\lambda_{j}^{2}\right)}\right] Y^{*}_{\ell m}(\theta,\phi)  \label{eqn:rij1}
\end{equation}
\begin{equation}
s^{ij}_{\ell m} = \int d\Omega \left[ \alpha(\theta,\phi) \times U^{j}(\theta,\phi) \right] Y^{*}_{\ell m}(\theta,\phi)  \label{eqn:kij1}
\end{equation}
Assuming each map to be a sum of signal and noise components: $Q^{i} = Q^{i}_{0}+\delta Q^{i}$, $U^{i} = U^{i}_{0}+\delta U^{i}$ and  $\alpha = \alpha_{0}+\delta$, and equations~\ref{eqn:rij1}~and~\ref{eqn:kij1} become
\begin{equation}
q^{ij}_{\ell m} = \int d\Omega \left[  \frac{ Q^{i}_{0}+\delta Q^{i}  - Q^{j}_{0} - \delta Q^{j} }{2 \left( \lambda_{i}^{2}-\lambda_{j}^{2} \right) }\right] Y^{*}_{\ell m}(\theta,\phi)  \label{eqn:rij2}
\end{equation}
\begin{equation}
s^{ij}_{\ell m} = \int d\Omega \left[ (\alpha_{0} + \delta \alpha) \times (U^{j}_{0} +\delta U^{j}) \right] Y^{*}_{\ell m}(\theta,\phi)  \label{eqn:kij2}
\end{equation}
Then we separate terms that rely on any noise component
from the purely signal components and square in Fourier space.
Converting the integrals to summations over pixels of size $\Omega_{pix}$, and writing the uncertainty per pixel in a map $X$ as $\sigma_{X_{n}}$, the noise power spectra can be written as

\begin{table}
\centering
\begin{tabular}{l c c c}
Instrument & \pbox{9.5cm}{\quad $\nu$ \\ (GHz)} & \pbox{9.5cm}{FWHM\\ (acrmin)} & \pbox{10cm}{Noise Depth\\ ($\mu K$-arcmin)} \\
\hline \hline
QUIET  & 43.1 & 27.3 & 36  \\
       & 94.5 & 11.7 & 36  \\ \hline
\pbox{1cm}{\ \\ BICEP/\\KECK} & 95 & 30 & 3.4\\
       & 150 & 30 & 3.4\\   \hline
ACT  & 30 & 5 & 14 \\
             & 40 & 5 & 14 \\
             & 90 & 2.2 & 11 \\
             & 150 & 1.3 & 10\\
             & 230 & 0.9 & 35 \\ \hline
PLANCK & 30   & 33   & 210 \\
       & 44   & 24   & 240 \\
       & 70   & 13   & 300 \\ \hline
SPIDER & 90   & 49   & 15 \\
	   & 150  & 30   & 11\\
	   & 250  & 17   & 36\\ \hline
CMB-S4 & 40 & 4 & 1 \\
	   & 90 & 2 & 1 \\
	   & 150 & 1 & 1 \\
	   & 220 & 0.7 & 1\\
\end{tabular}
\caption[Quiet Simulation Paramters]{\label{tab:simulation parameters} Parameters used in CMB simulations. All parameters derived from \cite{quiet_instrument,cmb-s4,spider:2013,act:2014,bicepbmode:2015,planck:2015i}.}
\end{table}

\begin{equation}
N_{\ell}^{\Delta Q} = \sum_{n}^{N_{pix}} \frac{\sigma_{Q_{n}^{i}}^{2} -2\sigma_{Q_{n}^{i}}\sigma_{Q_{n}^{j}} + \sigma_{Q_{n}^{j}}^{2}}{ \left( 2(\lambda_{i}^{2}-\lambda_{j}^{2}) \right)^{2}} \frac{\Omega_{pix}^{2}}{4\pi}  \label{eqn:noise-delta-q}
\end{equation}
\begin{equation}
N_{\ell}^{\alpha U} =  \sum_{n}^{N_{pix}} \left[ \sigma_{\alpha} U^{j}_{0} + \alpha_{0}\sigma_{ U^{j}} + \sigma_{\alpha} \sigma_{ U^{j}} \right]^{2} \frac{\Omega_{pix}^{2}}{4\pi}  \label{eqn:noise-alpha-u}
\end{equation}
Inserting these into equation~\ref{eqn:cross-correlation-variance}, a theoretical estimation of the variance in the cross-correlation power spectrum can be computed. This process can also be repeated to estimate the noise for the correlations of equations~\ref{eqn:dualphaalm}~and~\ref{eqn:qalm}. The two separate variance estimators are then added in quadrature to provide an estimate of the uncertainty for \autoref{eqn:correlator}. We also produce noise estimators by performing Monte Carlo simulations over independent noise simulations. The agreement between these estimators is shown in \autoref{fig:correlations}.  We find a good agreement between these two estimators in general. Variations between them can result from a spatial structure that exists in the noise maps not captured by an RMS thermal noise power.
}

}

\begin{figure*}
\centering
\includegraphics[width=.4\textwidth]{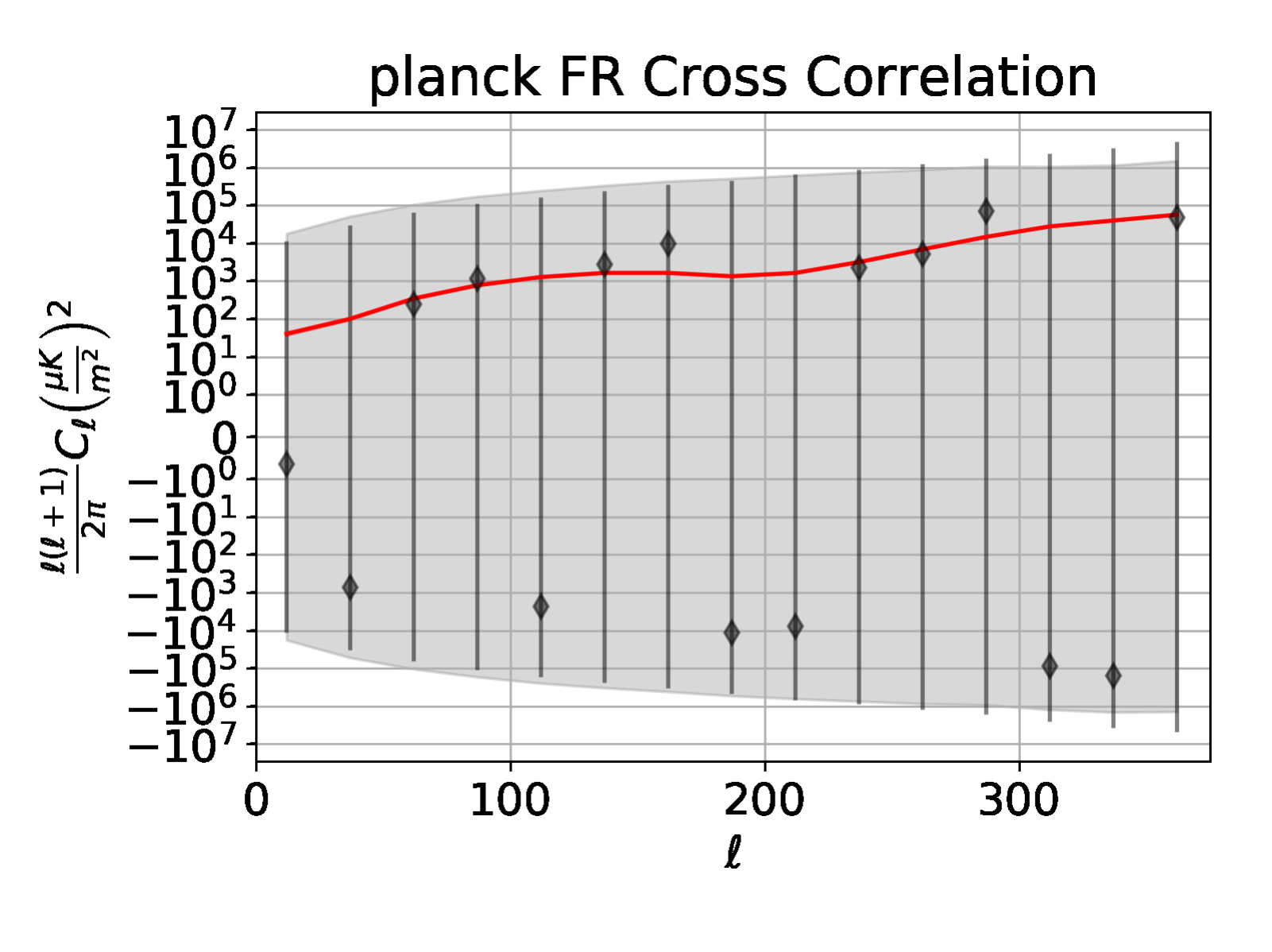}
\includegraphics[width=.4\textwidth]{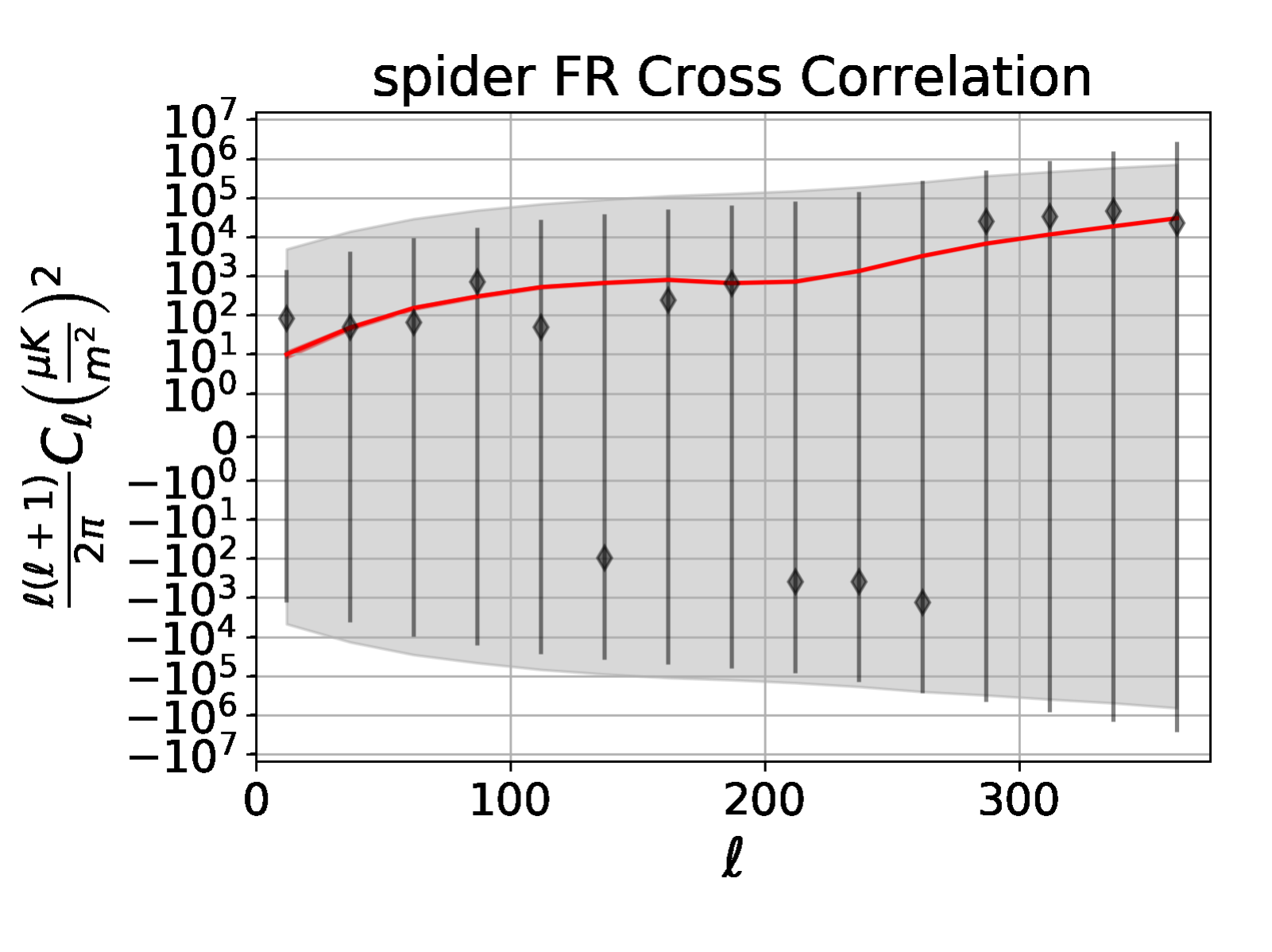}

\includegraphics[width=.4\textwidth]{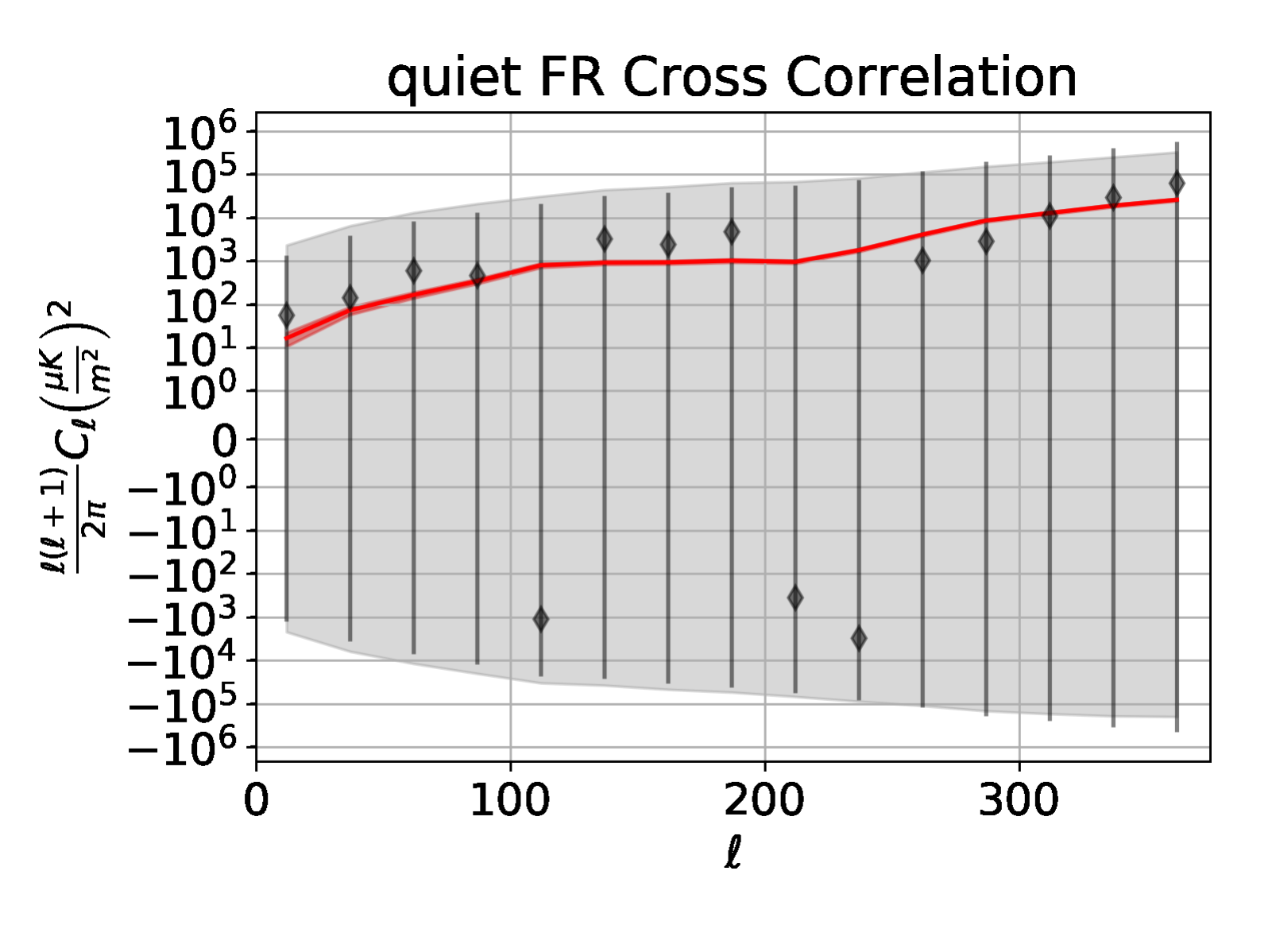}
\includegraphics[width=.4\textwidth]{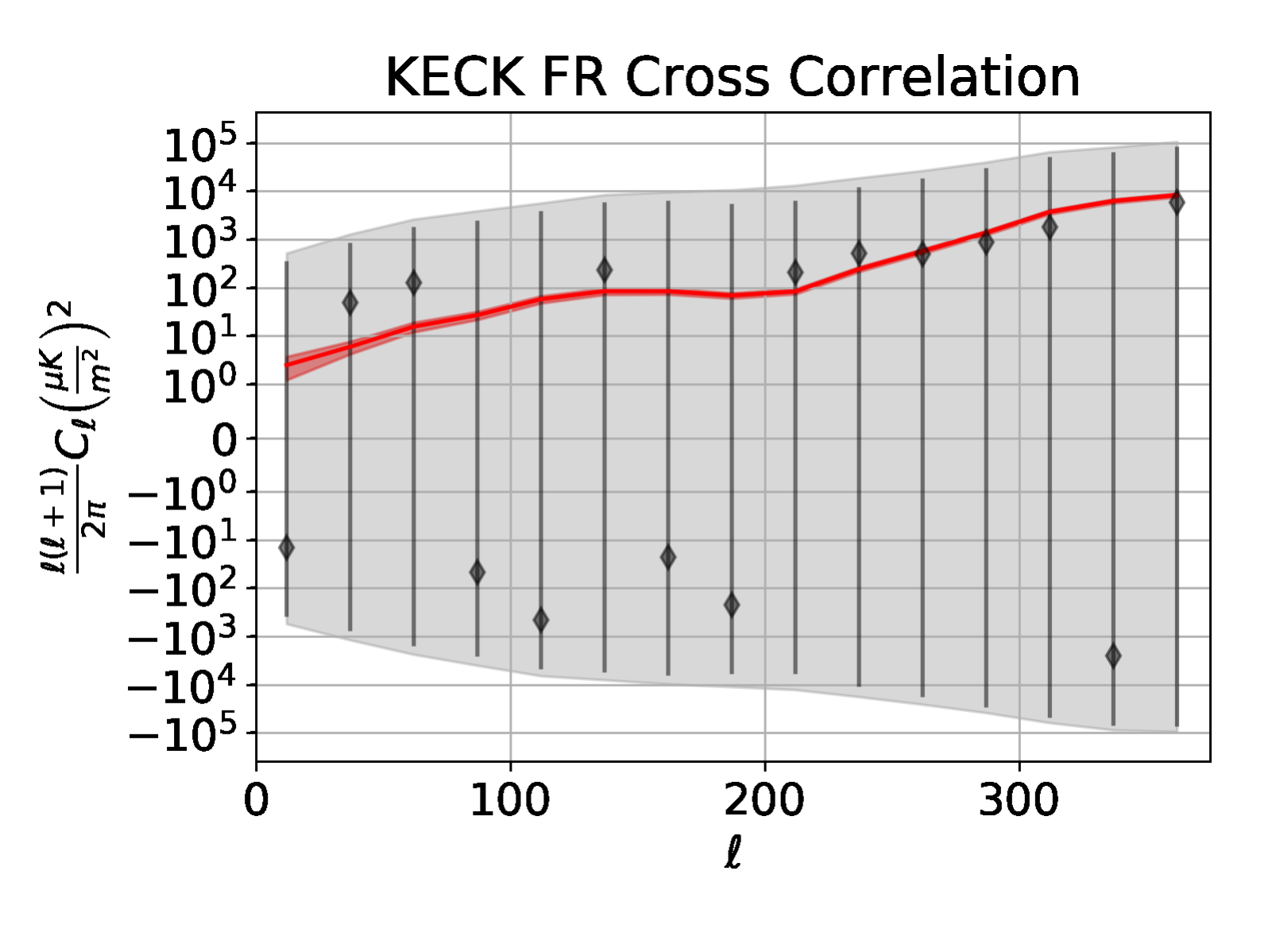}

\includegraphics[width=.4\textwidth]{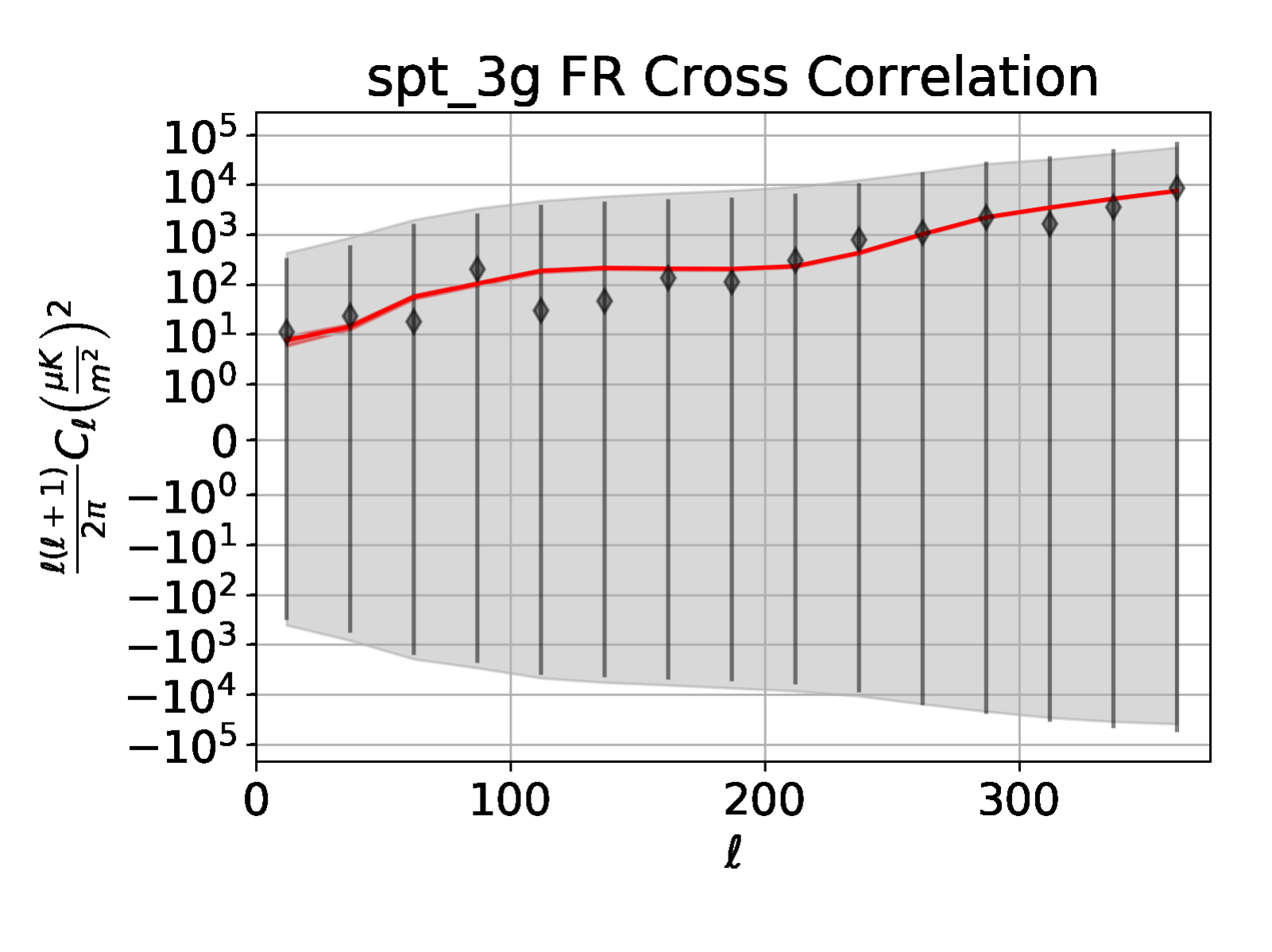}
\includegraphics[width=.4\textwidth]{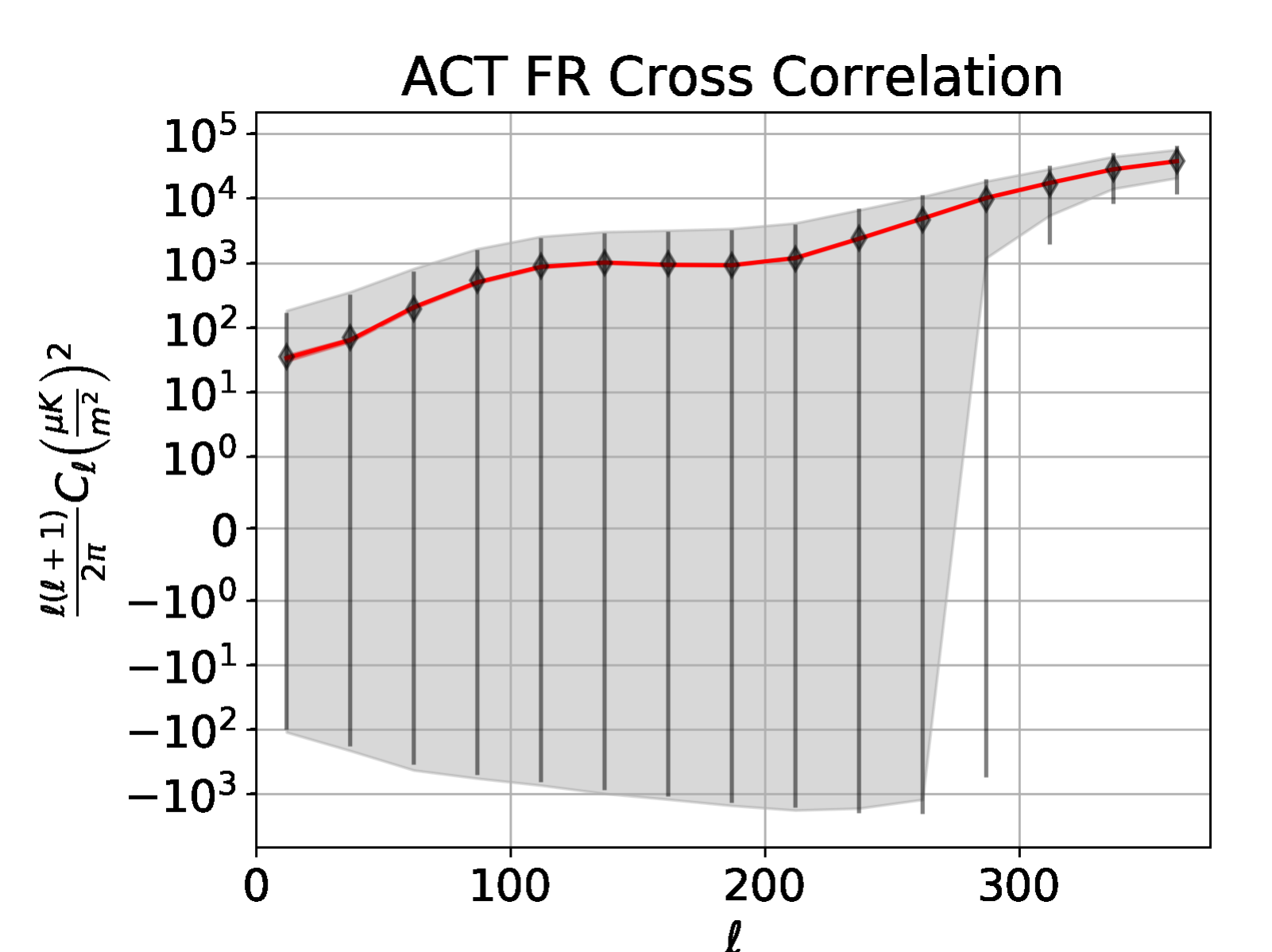}

\includegraphics[width=.4\textwidth]{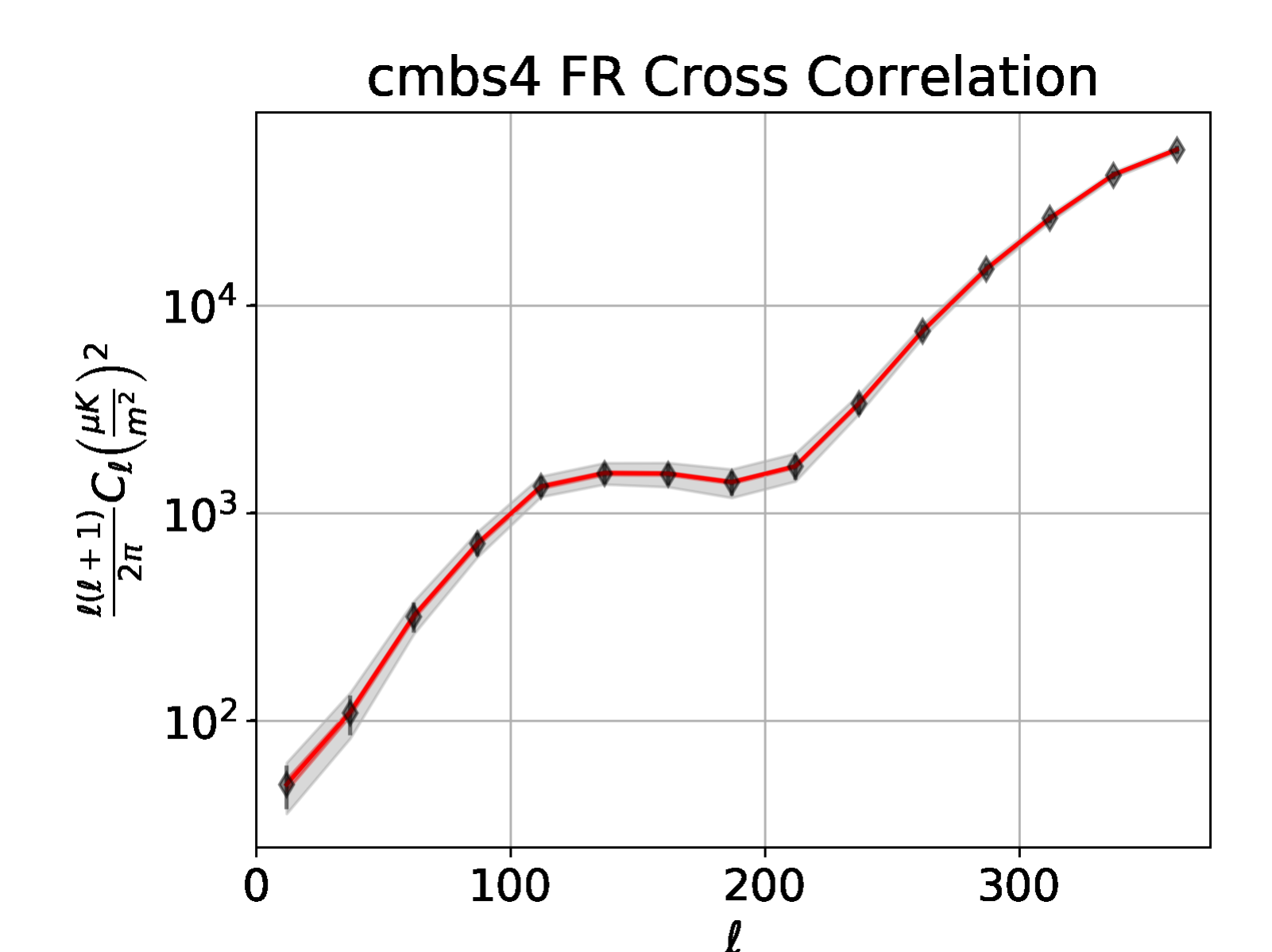}
\caption{Simulated output power spectra from QUITE, PLANCK, SPIDER, SPT-3G, AdvancedACT, and CMB-S4. The grey region represents the noise estimator given in \autoref{eqn:cross-correlation-variance}, error bars are generated by Monte Carlo trials over $N=500$ noise simulations. The theoretical and MC error bars show agreement within the simulated regions. The red line and surrounding red shaded region represent the theoretical signal for each experiment and uncertainty due to cosmic variance. Large thermal noise in current generation CMB arrays and limited spatial resolution of the FR map limit the signal to noise of this correlation. Experiments like the future CMB-S4 will have sufficient sensitivity to make high signal-to-noise detections of this signal. Signal-to-noise for each experiment is computed for $25<\ell<250$ due to the limited resolution of the radio RM maps. Higher resolution maps would allow for experiments like AdvancedAct to make detections at higher multipole moments.
\label{fig:correlations}}
\end{figure*}

\begin{figure*}
\centering
\includegraphics[width=.4\textwidth]{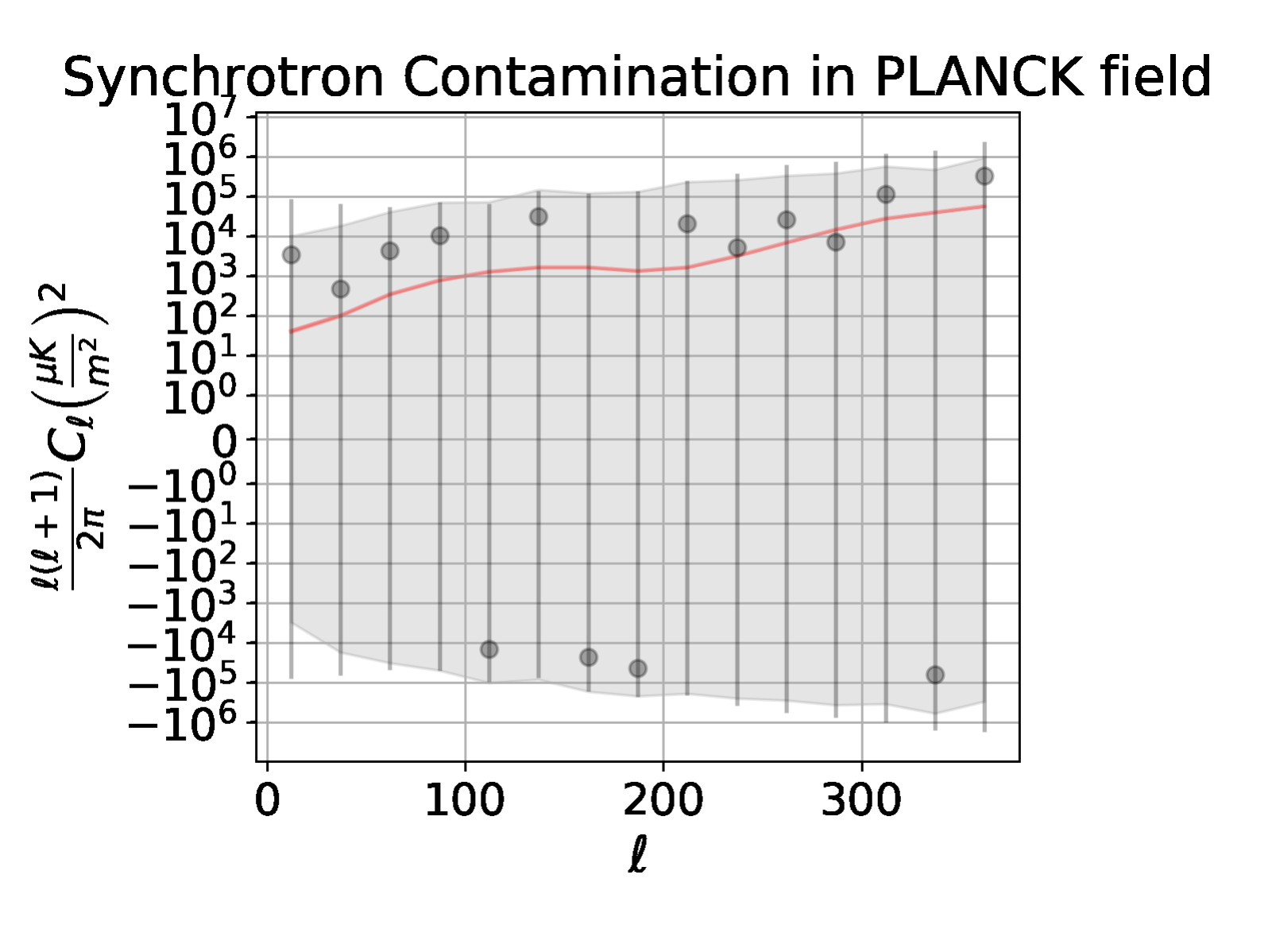}
\includegraphics[width=.4\textwidth]{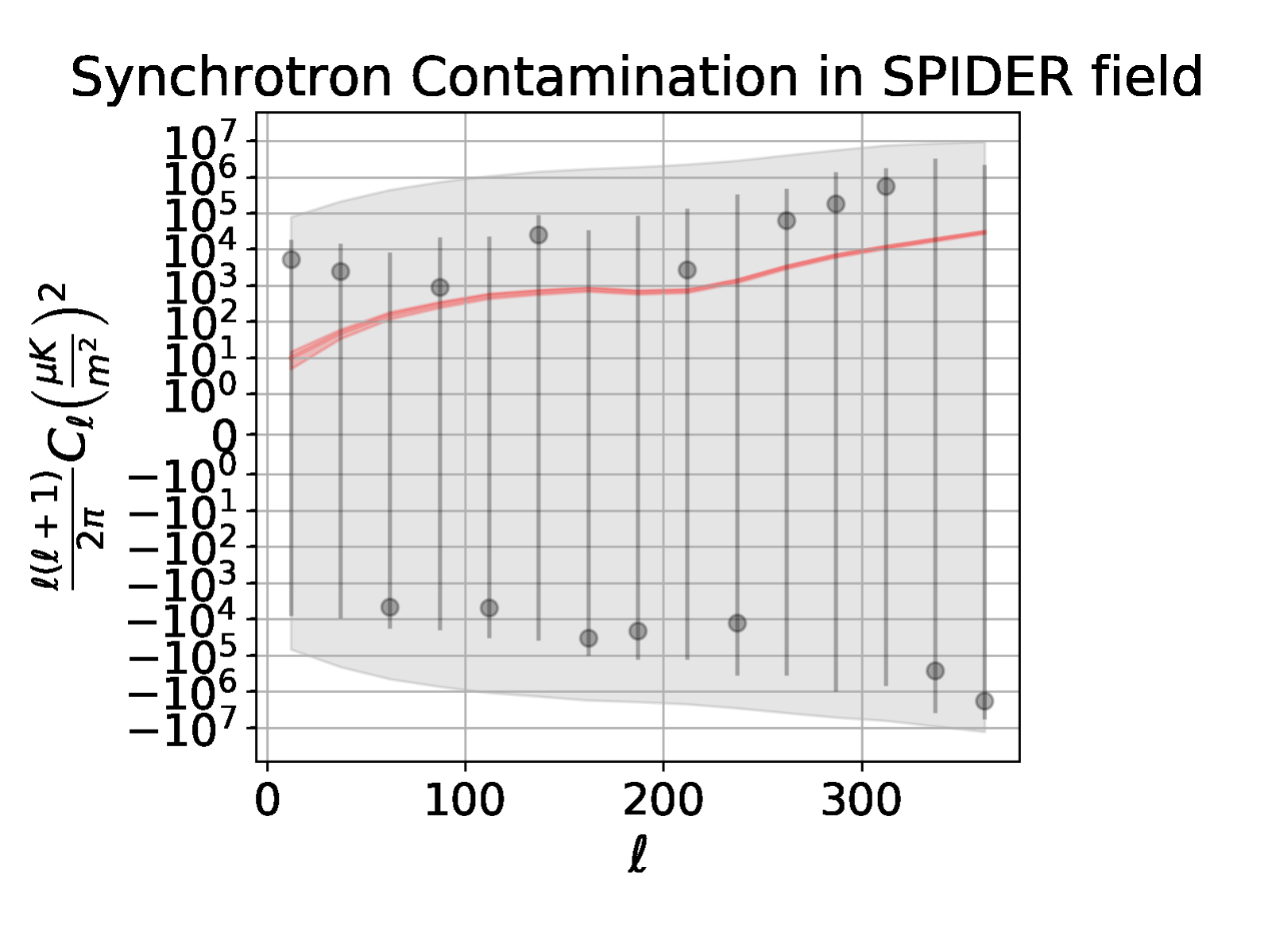}

\includegraphics[width=.4\textwidth]{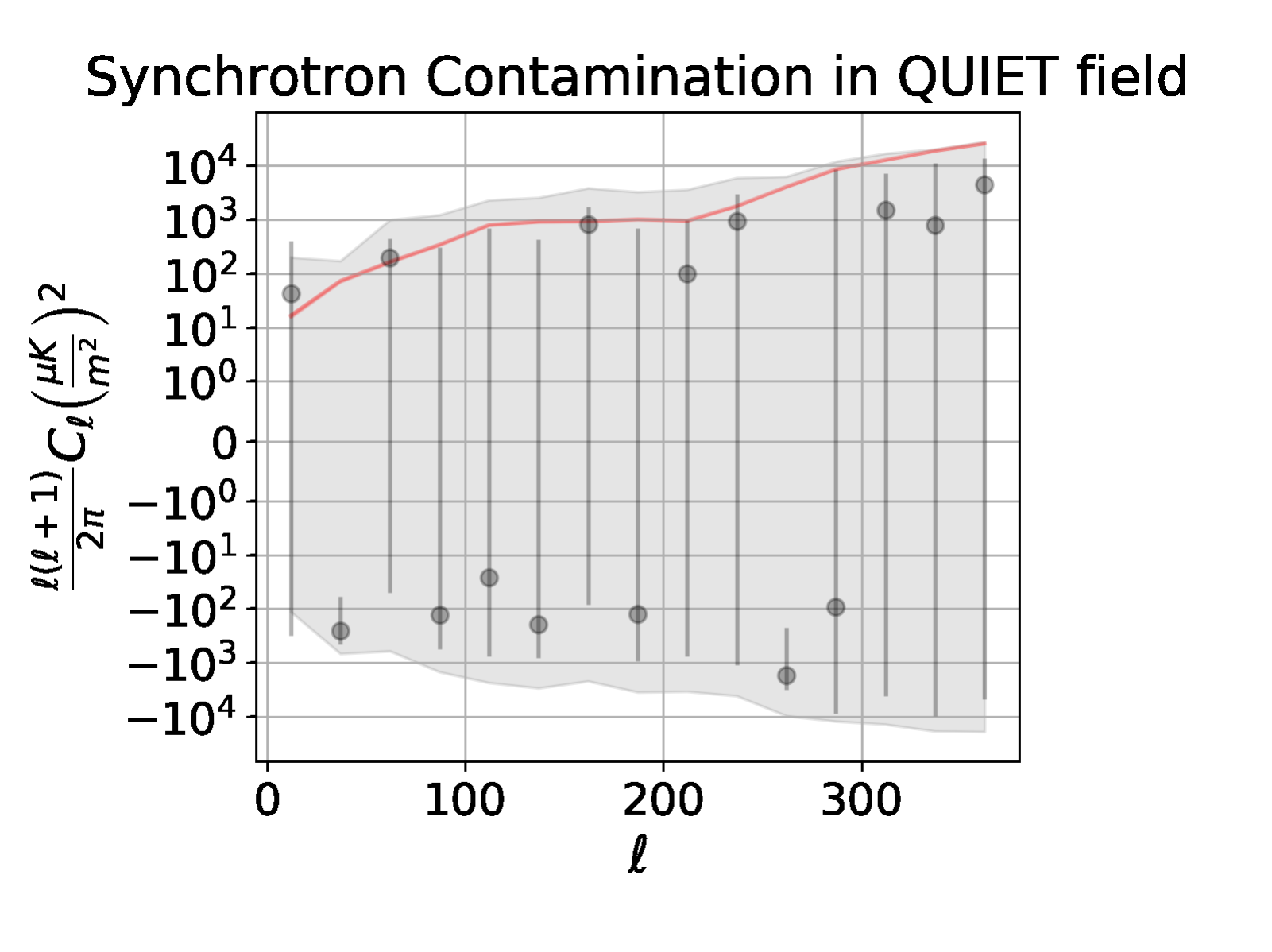}
\includegraphics[width=.4\textwidth]{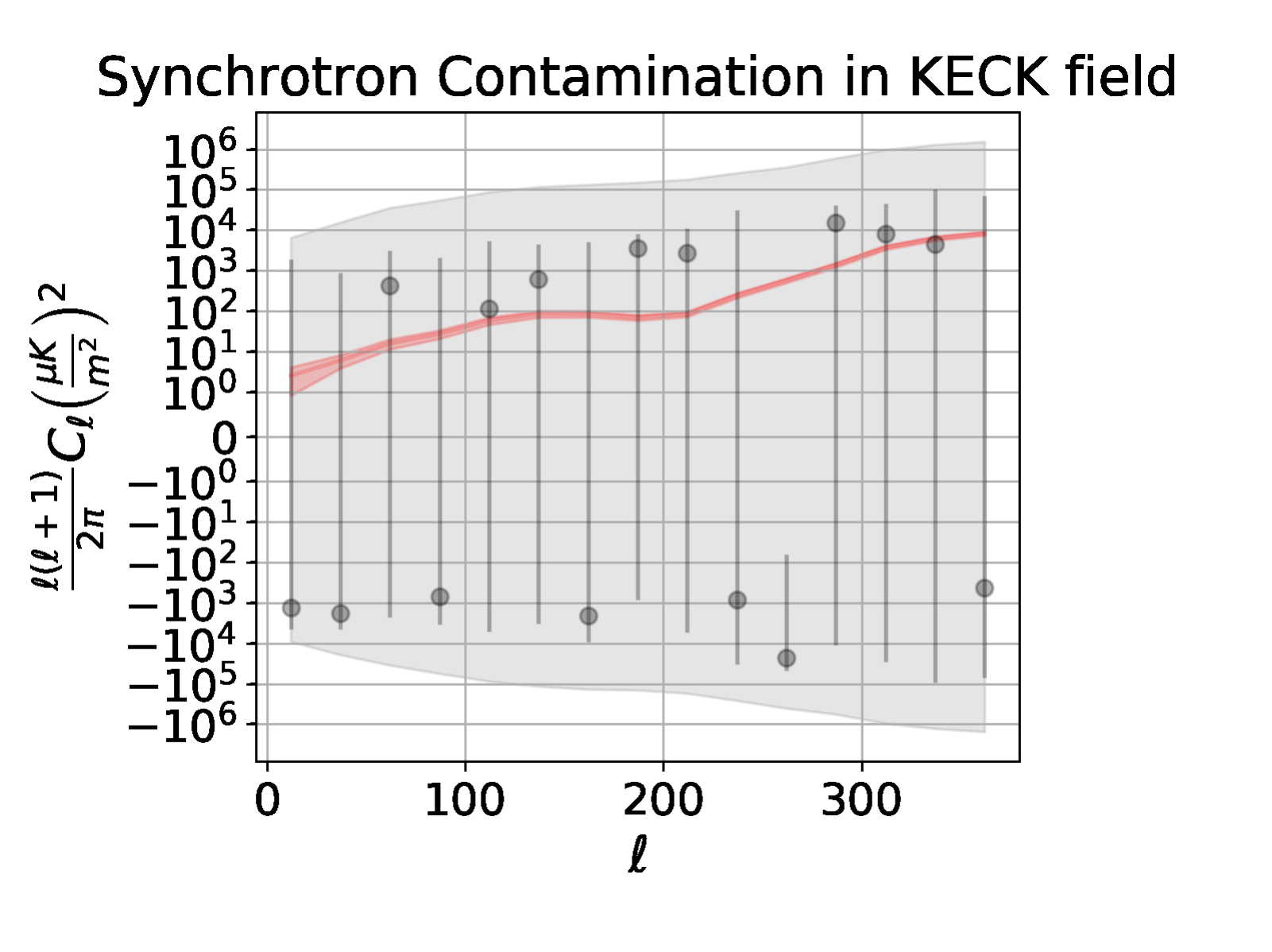}

\includegraphics[width=.4\textwidth]{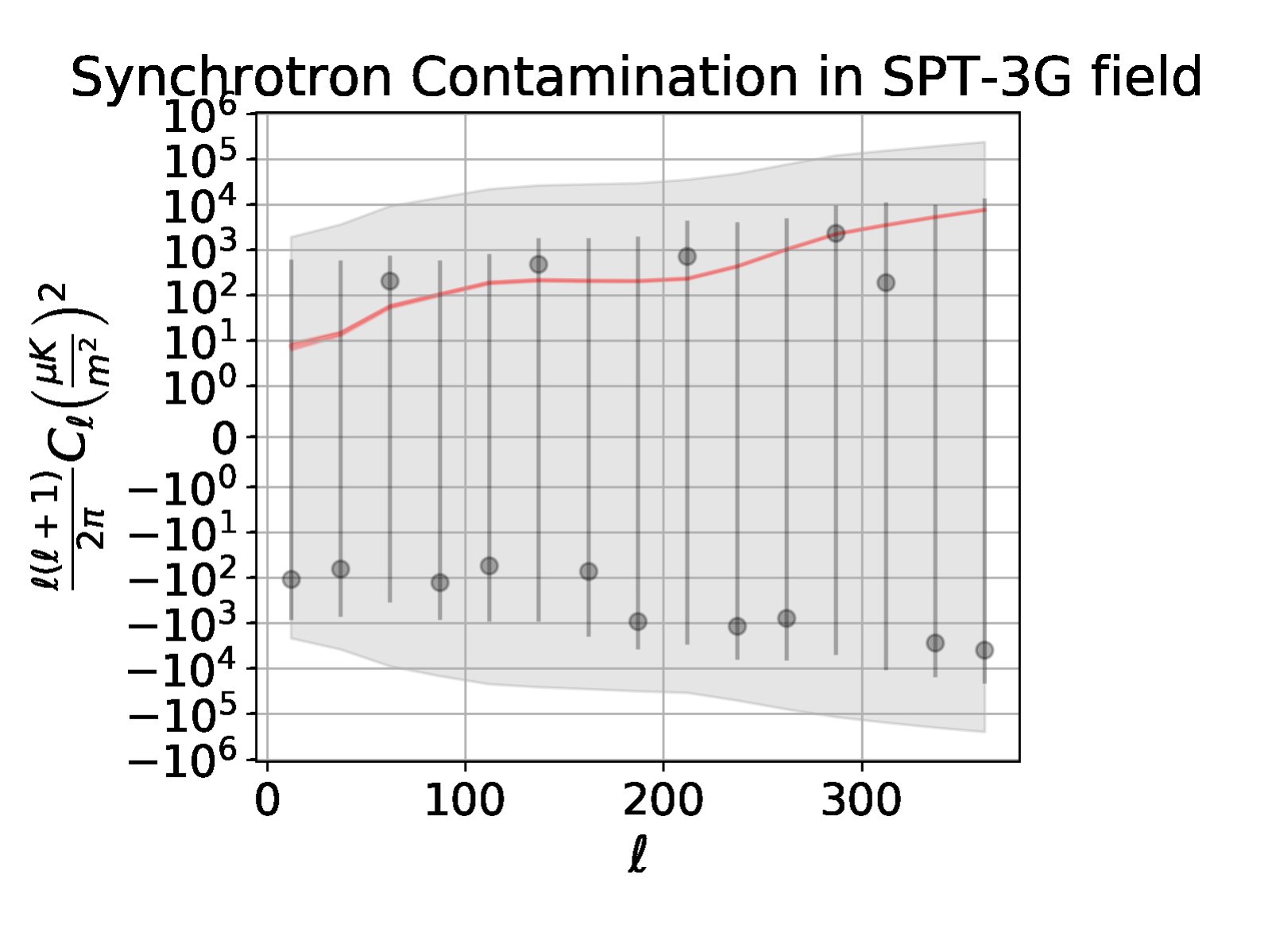}
\includegraphics[width=.4\textwidth]{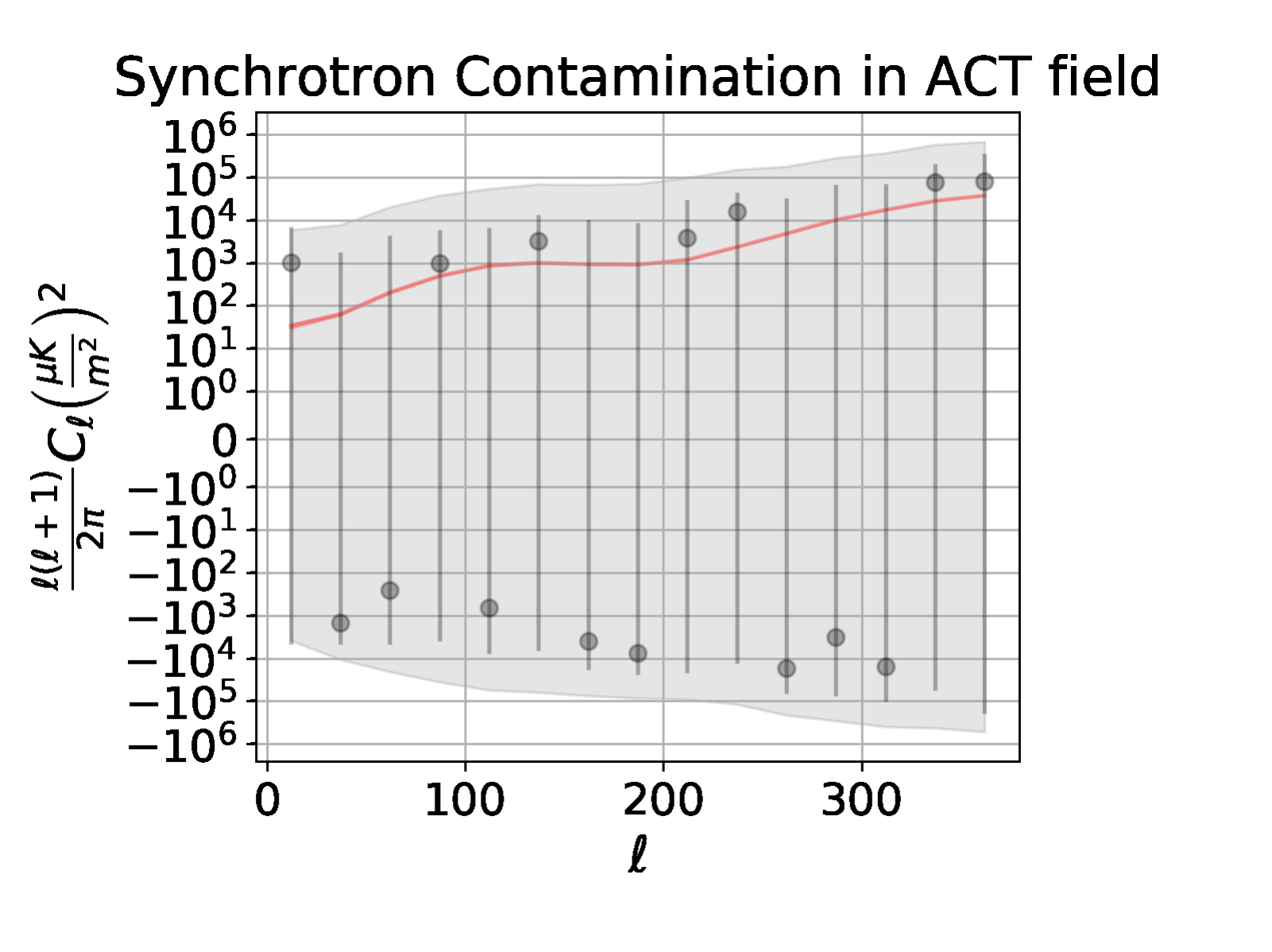}

\includegraphics[width=.4\textwidth]{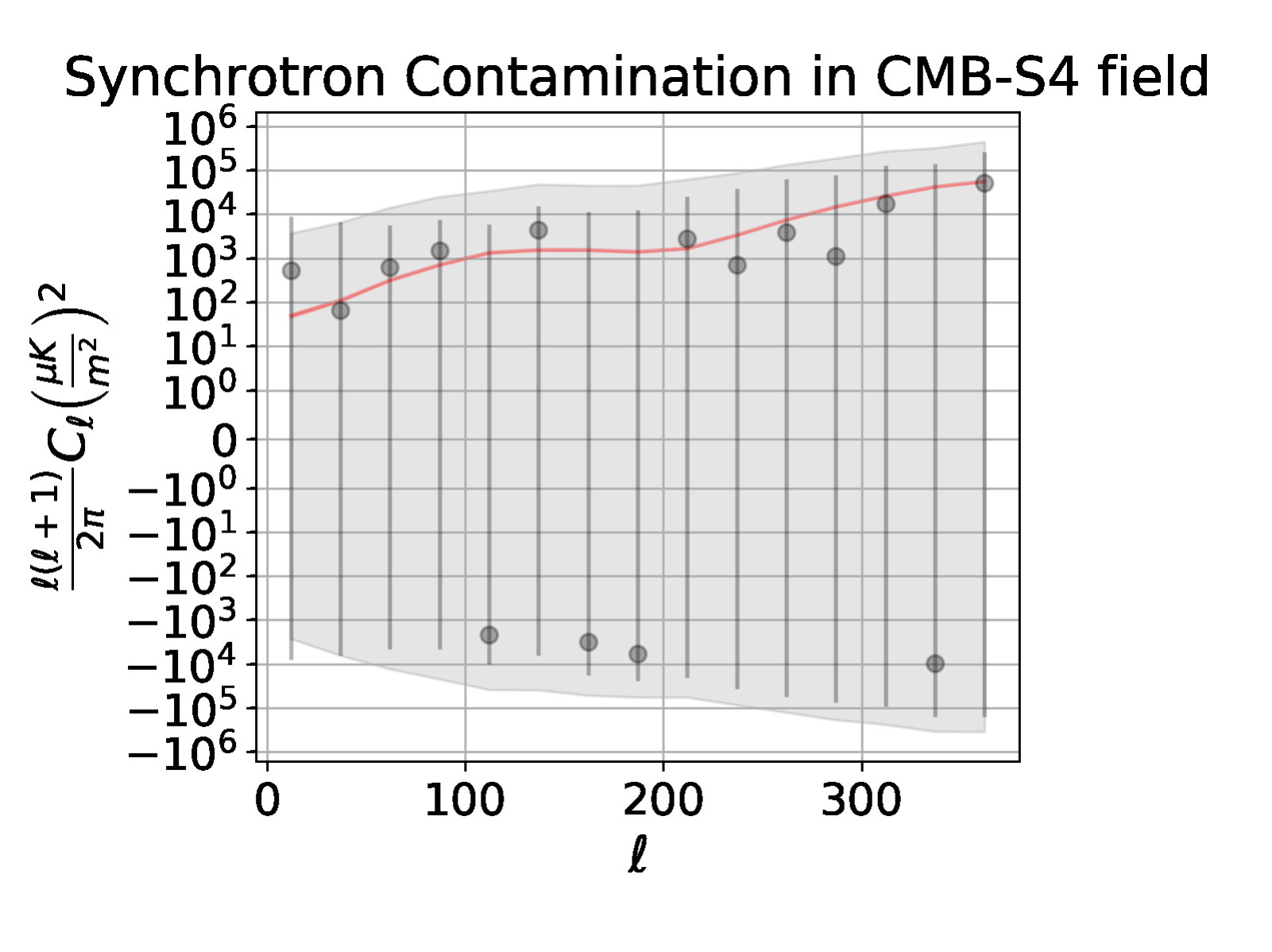}
\caption{Same as \autoref{fig:correlations}, but for inputs maps containing only polarized synchrotron. The grey shaded region is the theoretical error bar with the jack-knife error. Error bars represent the standard deviation in a bin. The red line is the theoretical estimate of the FR cross-correlation signal in the region. Synchrotron emission can contribute power on the same scale or higher as the expected FR signal. This power is dominated by the uncertainty in the correlation however. Good synchrotron removal is necessary to perform this correlation properly. The construction of synchrotron templates allows for high-precision subtraction to be performed on CMB data.
\label{fig:frxfore_sync}}
\end{figure*}

\begin{figure*}
\centering
\includegraphics[width=.4\textwidth]{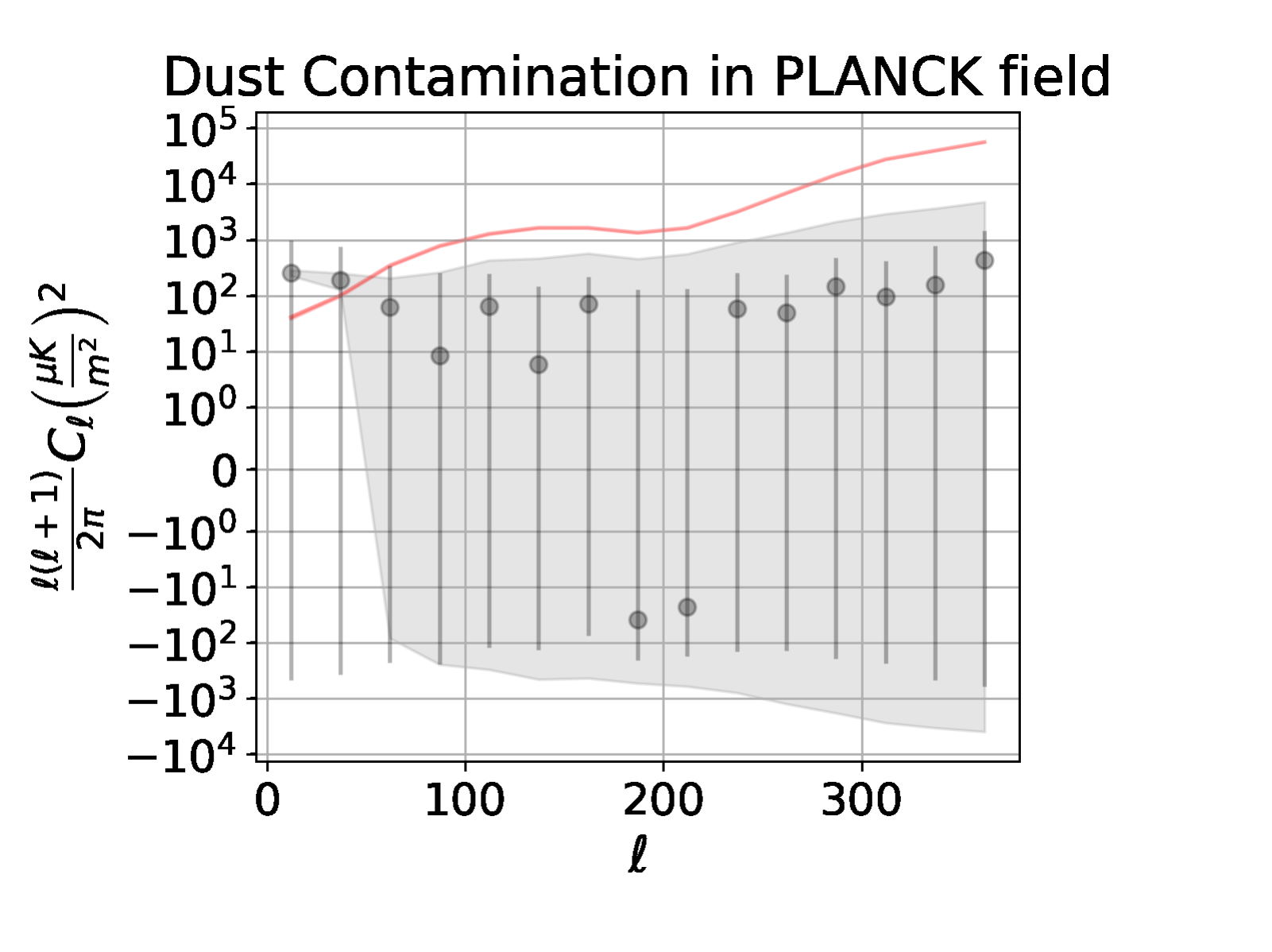}
\includegraphics[width=.4\textwidth]{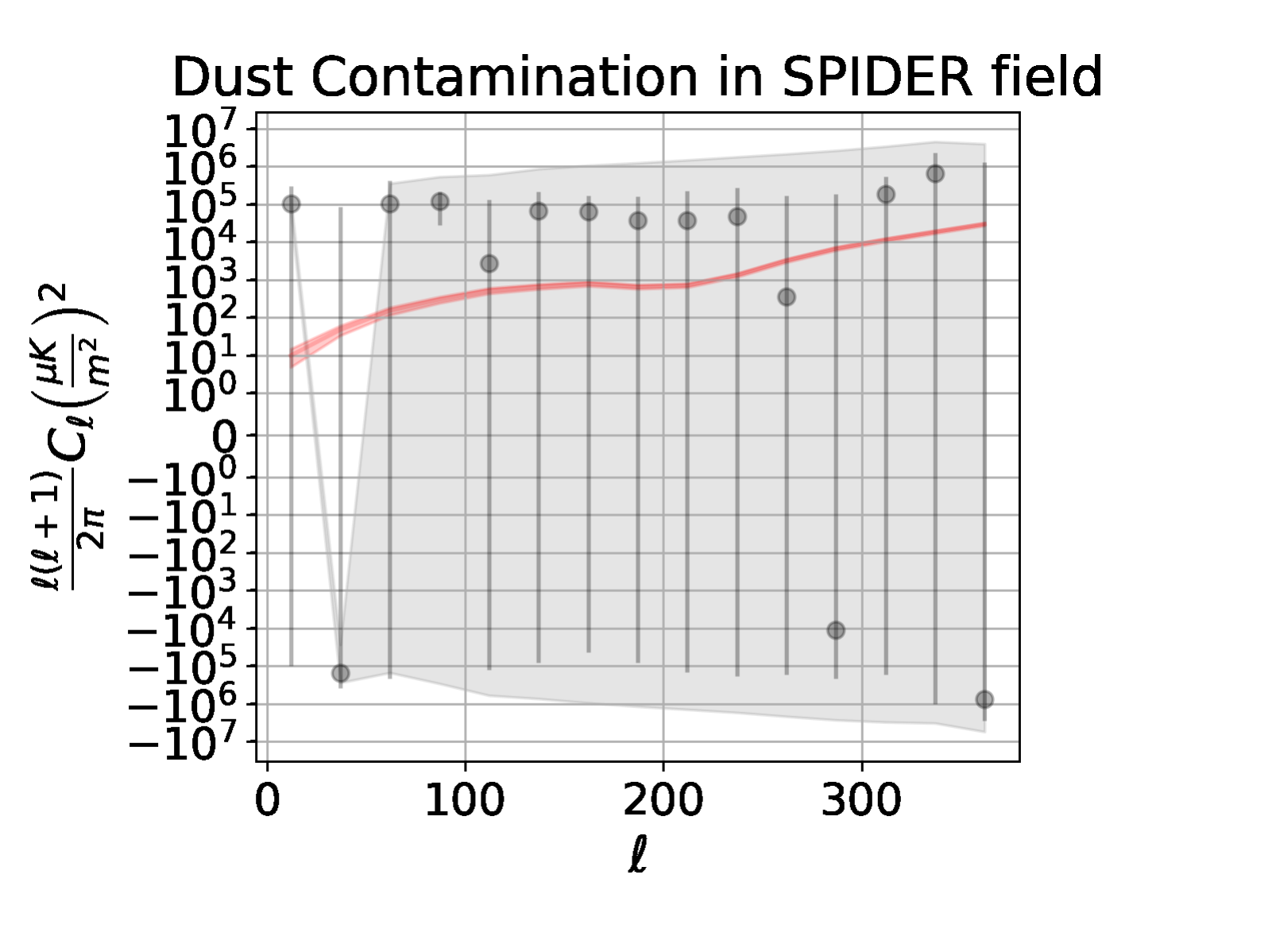}

\includegraphics[width=.4\textwidth]{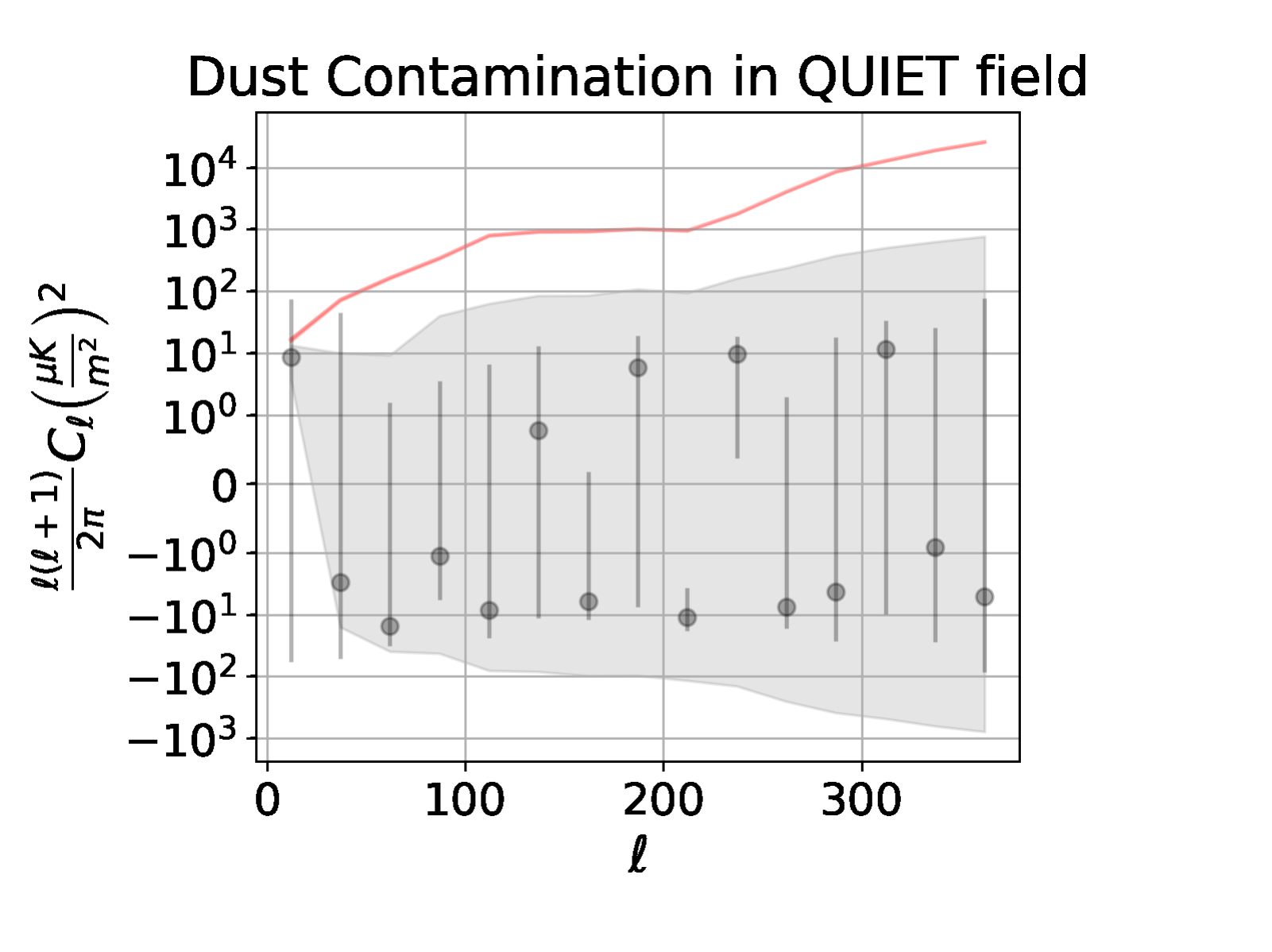}
\includegraphics[width=.4\textwidth]{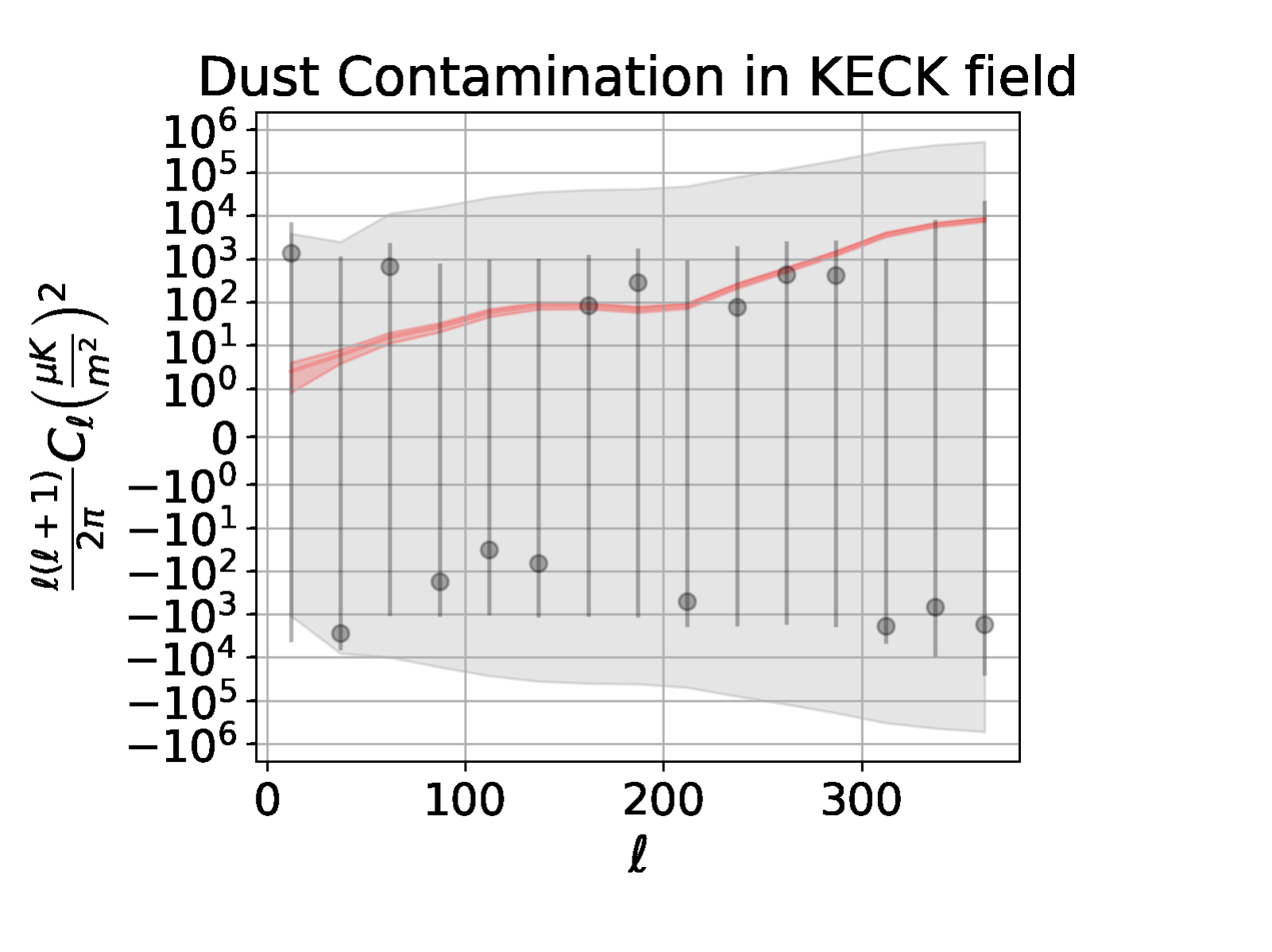}

\includegraphics[width=.4\textwidth]{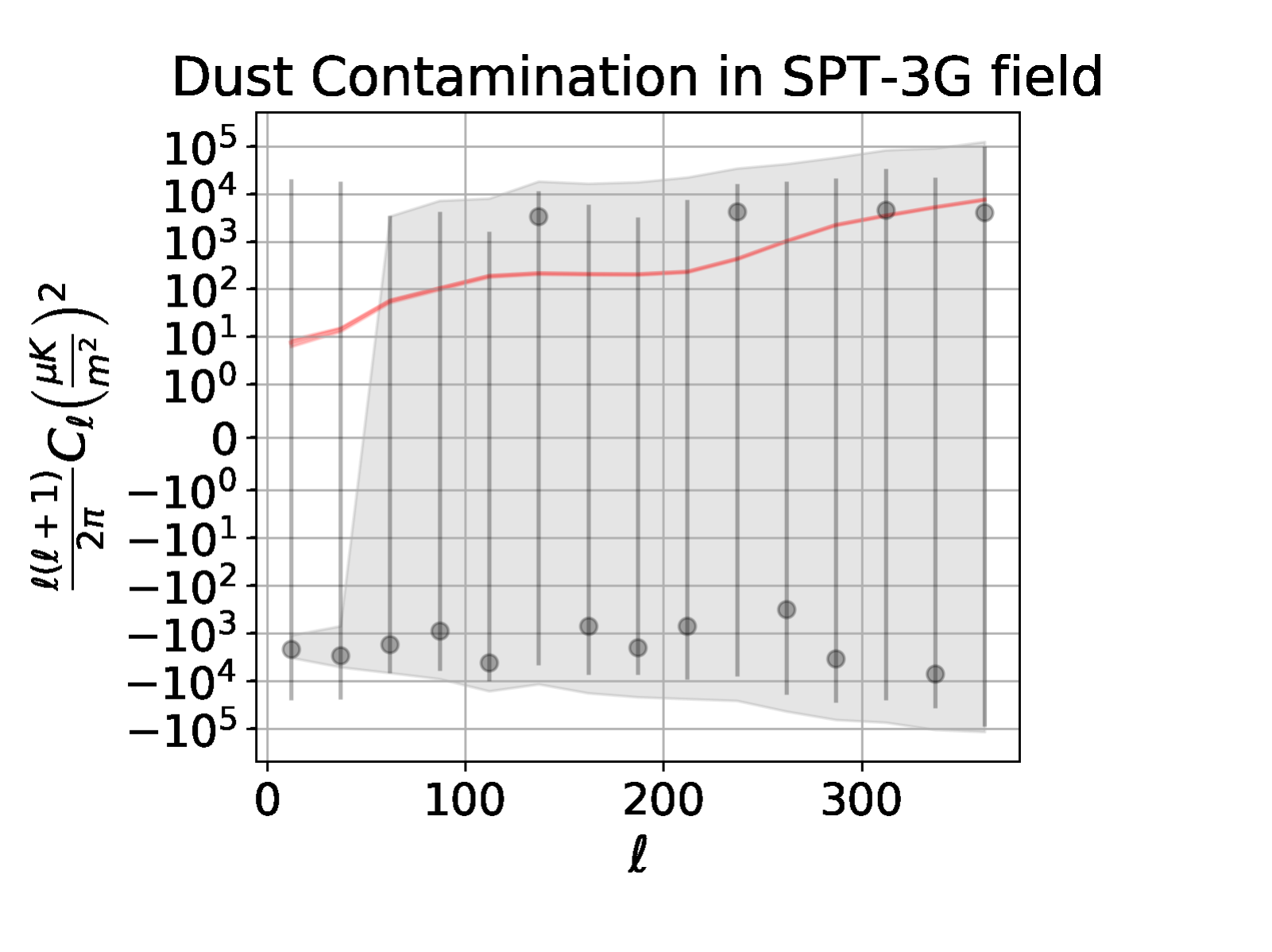}
\includegraphics[width=.4\textwidth]{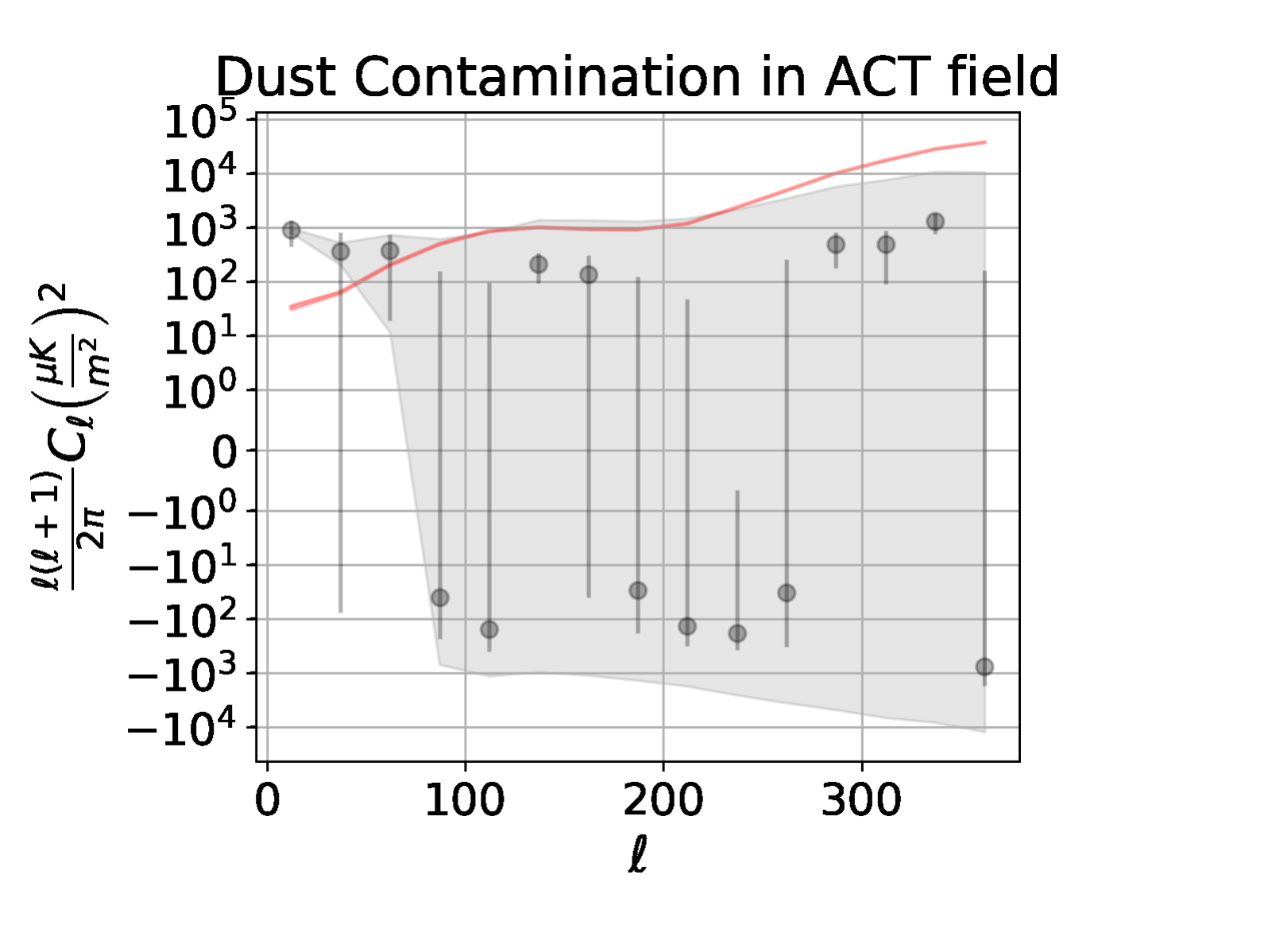}

\includegraphics[width=.4\textwidth]{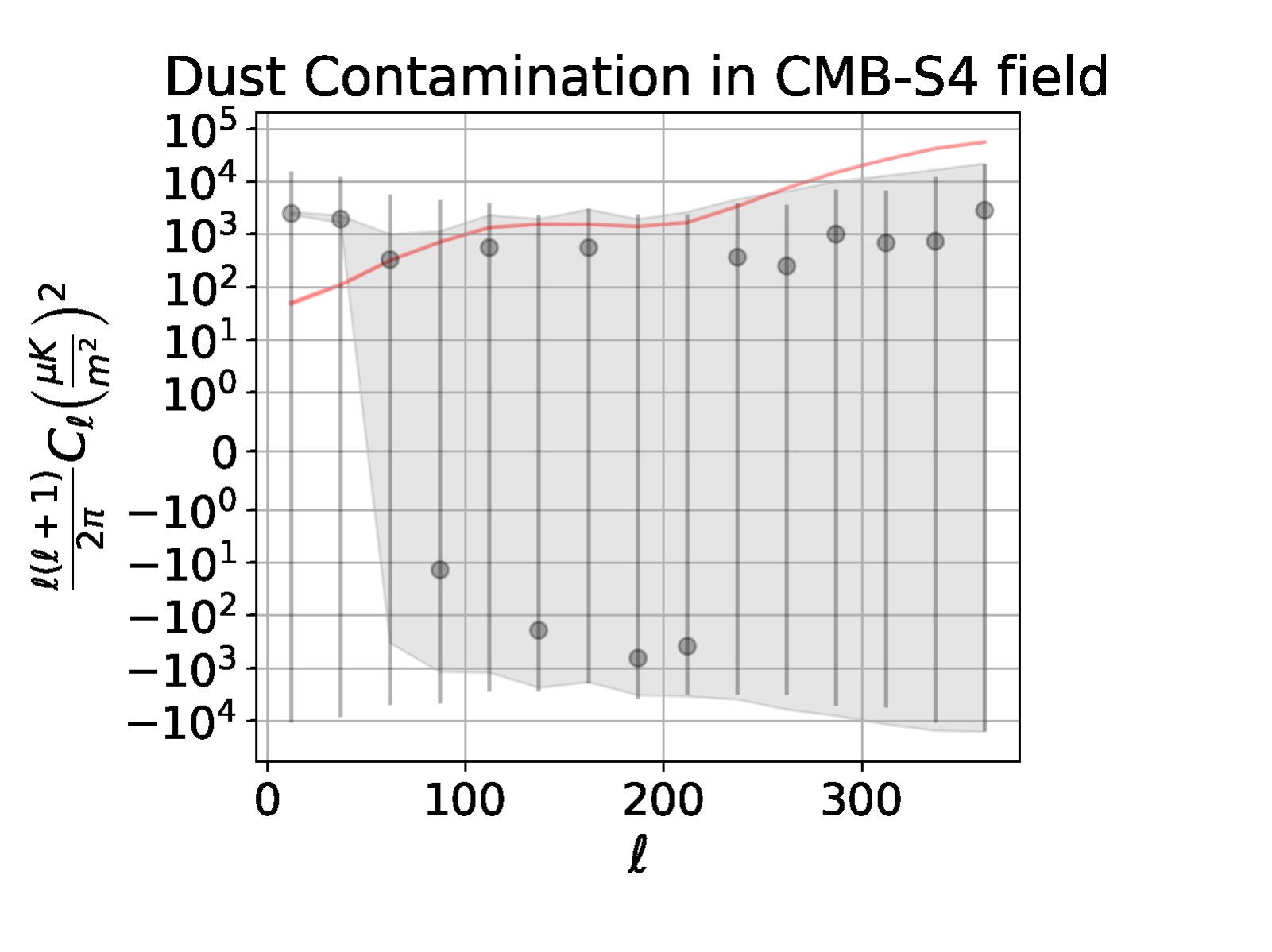}
\caption{Same as \autoref{fig:correlations}, but for inputs maps containing only polarized dust from Planck. The grey shaded region is the theoretical error bar with the jack-knife error. Error bars represent the standard deviation in a bin. The red line is the theoretical estimate of the FR cross-correlation signal in the region. Some experiments observe in regions with low dust and are not subject to this foreground. Experiments with high dust power also exhibit large uncertainty, and this would contribute to the uncertainty in the correlator. Overall, large uncertainty in the correlation requires good removal of polarized dust emission. \label{fig:frxfore_dust}}
\end{figure*}

\section{Simulation}\label{sec:simulation}{
We simulate CMB observations for the following surveys: QUIET, PLANCK, BICEP/KECK, AdvancedACT, PLANCK, SPIDER and sensitivities expected for CMB-S4. The parameters used to construct these simulations are described in \autoref{tab:simulation parameters}.
The {\scriptsize CAMB} software is used in simulating CMB data \citep{Lewis:1999bs, Howlett:2012mh} and the {\scriptsize HEALPIX}\footnote{Information on HEALPix available at http://healpix.sf.net/} \citep{gorski_et_al2005} and {\scriptsize ANAFAST} packages are used in data processing.

We generate pure CMB simulated data and apply FR to the polarized $Q$ and $U$ maps based on the frequencies of a given array.
The RM used to facilitate FR in these simulations is the map provided by \cite{Oppermann:2014} and shown in \autoref{fig: Radio RM}.

The polarized $Q$ and $U$ maps are then smoothed to the observing 
resolution for each instrument.
White noise is added to the smoothed polarized maps before an additional smoothing with a Gaussian beam to the desired resolution for analysis.
Noise is added before the final smoothing since data product maps are created at different spatial resolutions but must be smoothed 
to the same resolution for before this analysis. 
In order to properly difference maps of $Q$ and $U$ from different 
observational wavelengths, we smooth all maps to resolution of
$40$arcmin. Since the FR maps used in the analysis are 
only available at Healpix $NSIDE=128$, we downgrade all maps
to this Healpix $NSIDE$ and smooth to avoid pixelization effects.

 These images are then analysed using the method from \autoref{sec:correlation detection}. The results from these simulations are characterized using the following definitions.

To find the likelihood of a correlation, we calculate the posterior distribution for a scalefactor $\beta$ such that

\begin{equation}
C_{\ell}^{FR} = \beta C_{\ell,th}^{FR}  \label{eqn:beta-cl}
\end{equation}
with the $C_{\ell,th}^{FR}$ given from noiseless simulations in the observed region. Due to the low angular resolution of current FR measure maps, this posterior is only fit for $25 < \ell \leq 250$. Assuming a uniform prior distribution for $\beta$ in the range [--1000,1000] and Gaussian variances, when computing the posterior distribution of the simulations, the standard error of the mean is used as the uncertainty of the data points. Defining the estimator of $\beta$ as $\bar{\beta}$, it should fall in the interval

\begin{equation}
\bar{\beta} = 1 \pm \frac{\sigma_{\bar{\beta}}}{\sqrt{N_{sim}}}
\end{equation}
where $\sigma_{\bar{\beta}}^{2}$ is the variance of the posterior distribution and $N_{sim}$ is the number of simulations.
We calculate the total SNR of the correlator as
\begin{equation}
SNR = \frac{1}{\sigma_{\bar{\beta}} }  \label{eqn:snr-1}
\end{equation}
 Since the simulations are designed such that $\beta =1$, a strong correlation is represented by a narrow peak centred around $\bar{\beta}=1$.  The posterior is computed in this way to determine whether there exists bias in the simulations or correlator.

The results from our simulations are illustrated in \autoref{fig:correlations}.
The signal to noise expected from this correlation can be found in \autoref{tab:snr-estimate}.
Based on our estimates, current CMB experiments are unable to detect the FR cross-correlation due to the high thermal noise in CMB observations and the limited spatial resolution of current RM maps.
 Next generation experiments with thermal noise levels similar to the CMB-S4 estimates and increased spatial resolution of RM maps will be able to make high signal-to-noise detections.
 }

\section{CMB Foregrounds}\label{sec:foregrounds}{

 The presence of polarized CMB foregrounds like synchrotron and thermal dust emission may cause false correlations with this method if present in polarization observations.
 
  An in-depth analysis and discussion of the characteristics of polarized CMB foregrounds can be found in \citet{planck:2016diffuse}. As noted by \citet{dineen:2004}, the intensity of synchrotron emission is a function of the energy density of electrons, $N(E)dE$, and the strength of magnetic field. When the electron energy density exhibits a power-law distribution

\begin{equation}
N(E)dE \propto E^{-(2\alpha +1)}dE
\end{equation}
the intensity of synchrotron radiation takes the form
\begin{equation}
I(\nu) \propto B_{\perp}^{1+\alpha}\nu^{-\alpha}
\end{equation}
where $B_{\perp}$ is the component of the magnetic field perpendicular to the line of sight  and $\alpha$ is the spectral index.

This provides information on the total synchrotron intensity, the 
polarized components of synchrotron emission are described in \citet{orlando_strong2013} as

\begin{equation}
Q \propto \int(B_{\perp x}^{2} - B_{\perp y}^{2}) I(s)d\mathbf{s}
\end{equation}

\begin{equation}
U \propto \int(2B_{\perp x}B_{\perp y}) I(s)d\mathbf{s}
\end{equation}
where the $B_{\perp x}$ and $B_{\perp y}$ are the components of the 
magnetic field perpendicular to the line of sight and the integral
is performed along the line of sight.

While synchrotron emission and FR are dominated by components of the 
same magnetic field, $B$, some field configurations may produce
the existence of one effect and not the other (e.g. FR without synchrotron and vice-versa). 

Polarized thermal dust emission is dominated primarily by the temperature of the dust, $T_{d}$, as well as the 
component of the magnetic field perpendicular to the line of 
sight $\mathbf{B_{\perp}}$ and its alignment with the orientation
of dust particles \citep{planck:2015xx}. 
While its correlation with FR cannot necessarily be intuitively
predicted, much work has been done attempting to simulate this
emission \citep{planck:2015xx,ghosh:2017,vansyngel:2017} and 
accurate simulations of emission can be used to estimate the 
amount of correlation.

\citet{dineen:2004} use a Spearman rank-order correlation coefficient to investigate the amount of correlation between FR from radio point sources and CMB foreground maps. 
They find a stronger correlation between dust and FR than between synchrotron and FR, even outside the galactic centre.

For our analysis we are investigating how polarized 
foreground may cause a false correlation with our method. The residual 
correlations found in the Spearman rank-order analysis suggest that 
polarized CMB foregrounds may produce false positive correlation.

To determine the extent to which the FR cross-correlation may be contaminated by other low-frequency polarized CMB foregrounds, we compute the correlator from \autoref{sec:correlation detection}  with polarized dust and synchrotron maps provided by PLANCK \citep{planck:2015fore} for each set of observed frequencies in the simulated experiments.

We can estimate the error in the foreground correlation by sending a jack-knifed map, created from differencing the Half mission 1 and Half mission 2 images from each potential foreground respectively, through the correlation pipeline. This error is represented by the shaded regions in Figs~\ref{fig:frxfore_sync}~and~\ref{fig:frxfore_dust}.

 Fig.~\ref{fig:frxfore_sync} shows the results of cross-correlation pipeline when the polarized synchrotron maps are substituted for rotated CMB images. Within error, this correlation is found to be consistent with zero and is not expected to contaminate a net signal to the correlation. The large magnitude of recovered correlations is comparable to the expected FR signal and indicates
that proper foreground removal from observation is necessary to ensure synchrotron contamination does not dominate the final error budget.


Similarly, Fig.~\ref{fig:frxfore_dust}  shows the results of the cross-correlation pipeline when the polarized dust maps are substituted for rotated CMB images. Polarized dust shows a residual correlation for the simulated frequencies. This signal is especially prominent in the high frequencies where dust emission is strong.
Hence, in the analysis of actual observations, care must be taken to accurately remove this foreground before performing the cross-correlation.

For both types of foregrounds analysed, the jack-knife errors dominate any potential residual signal. Without proper removal, these foregrounds may contribute to a false correlation or an anticorrelation. The existence of detailed foreground maps for direct subtraction and the usage of techniques like a principle component analysis allow for the removal of these foregrounds from CMB maps in practice.
}

\section{Application to Real Data}\label{sec:real-data}{
Based on the simulation, the low-frequency data collected by PLANCK should not be able to detect the effects of the Galactic FR. The contributions from thermal noise dominate the correlation.
In this section, we apply the correlator to actual PLANCK data to test the predictions above.

When applied to Planck LFI data, the cross-correlation produces the results
 shown in Fig.~\ref{fig:planck-data}. Black error bars in Fig.~\ref{fig:planck-data} are given by the standard deviation of the power spectrum in a bin of $\Delta \ell =25$ over $N=500$ realization of the noise covariance map provided in the PLANCK data release injected into a simulated CMB signal.
The grey shaded region is the theoretical error given in Section~\ref{sec:uncertainty} with the thermal noise level from \cite{planck:2015i}.
 The posterior distribution of $\beta$ is generated using the theoretical error bars for the PLANCK data. 
  The posterior distribution for $\beta$ produces $\bar{\beta}=27.91$ and $\sigma_{\bar{\beta}}=74.00$.

 The mean of the posterior is consistent with zero within error and the standard deviation of the posterior, $\sigma_{\bar{\beta}}$, and the SNR from the analysis of the PLANCK data agrees with the expected level from simulation within a factor of $\sim 15\%$. The disagreement may be a result of a spatial structure that exists in the noise covariance map provided by PLANCK opposed to the RMS noise amplitude used in simulations.
The agreement between the simulated and actual Planck analyses supports the predications made for other experiments as well.

}

\begin{table*}
\centering
\small
\begin{tabular}{l c c c c c c}
Experiment & \pbox{2.5cm}{Frequencies \\ \hspace*{.4cm} (GHz)} & \pbox{3cm}{Noise levels \\ ($\mu$K-arcmin)} & $f_{sky}$ & $\bar{\beta}$ & $\sigma_{\bar{\beta}}$ & \pbox{3cm}{\hspace{.5cm} S/N \\ Galactic FR }\\
\hline
PLANCK & 30,44,70 & 210,240,300 & 0.73$^{\bf \footnotesize 1.}$ & .21 & 63.80 & 0.02 \\
BICEP2 & 100,150 & 3.4 & 0.01 & 0.71 & 25.09 & 0.04 \\
SPIDER & 90,150 & 10  & 0.1 & 0.59 & 23.11 & 0.04\\
QUIET & 45,90 & 36 & 0.005 & 1.41 & 13.07 & 0.08\\ 
SPT3G & 150,220 & 3.5,6 & 0.06 & 0.90 & 8.42 & 0.11 \\
AdvancedACT & 30,40,90,150,230 & 14,14,11,10, 35 & 0.5 & 1.03 & 0.73 & 1.36\\
CMB-S4 & 40-220 & 1 & 0.73$^{\bf \footnotesize 1.}$ & 1.00 & 0.04 & 22.73 \\ \hline \hline
SPT3G & \multicolumn{3}{c}{\dotfill} & 0.88 & 4.13 & 0.24 \\
AdvancedACT &\multicolumn{3}{c}{\dotfill} &  1.03 & 0.35 & 2.89 \\
CMB-S4 & \multicolumn{3}{c}{\dotfill} &  1.01 & 0.02 & 45.46 

\end{tabular}
\caption[Estimated SNR]{Simulation results and estimates of the SNR expected using this method to detect Galactic FR in the CMB. $f_{sky}$ refers to total sky fraction observed. Model parameters taken from \cite{quiet_instrument,cmb-s4,spider:2013,act:2014,bicepbmode:2015,planck:2015i}. 
The lower rows of SPT3G, AdvancedACT and CMB-S4 represent the parameters of the fit if all available multipole moments are used in the correlation. 
This demonstrates that increasing fidelity at higher mulitpole moments of the input maps will allow for stronger detections.  \label{tab:snr-estimate} \\ {\footnotesize {\bf 1.} .73 sky fraction based on WMAP nine-year polarization analysis mask } }
\end{table*}

\section{Discussion}\label{sec:summary}{

\begin{figure}
\centering
\includegraphics[width=.45\textwidth]{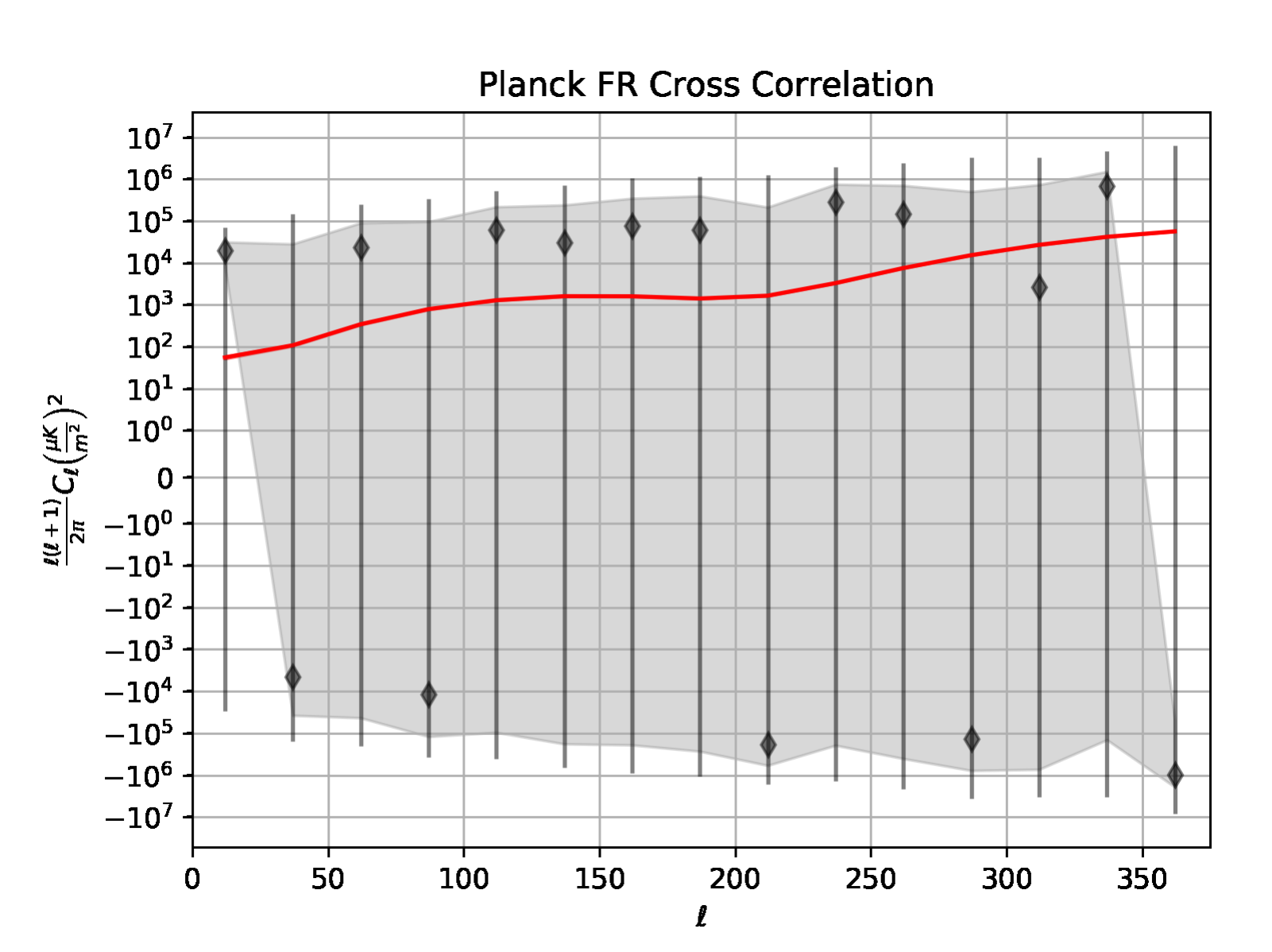}

\caption{Real Planck LFI data correlation, the inverse variance weighted sum of 30x44, 44x70 and 30x70 correlators.
The error bars are derived from the standard deviation within a bin of $\Delta \ell=25$. The correlator is consistent with zero for all tested multipoles, indicating no observable signal in the data. The grey shaded region represents the theoretical error obtained obtained from the covariance provided in the planck data. Discrepancy between the error bars and shaded regions may be a representative of the limited statistics in the data.\label{fig:planck-data}}
\end{figure}  

The simulations and analysis of this cross-correlation method for detecting FR in the above sections have addressed two types of surveys. The all-sky survey, which provides large sky coverage, and the single field survey, which is limited in sky coverage but can integrate to lower noise levels.

 The simulation results and estimates of the expected SNR for various survey configurations are displayed in Table~\ref{tab:snr-estimate}.
The analysis of these simulations is conducted for $25 < \ell< 250$.
We predict a signal near $2\sigma$ significance in AdvancedACT data and a detectable signal at very high significance ($> 10\sigma$) in a future CMB-S4 experiment. 
If we fit for $25<\ell<384$ (the full multipole resolution of the RM map), we find an increase in statistical significance for SPT3G, AdvancedACT and the CMB-S4 experiments. This demonstrates that 
increasing spatial resolution of maps used in this analysis will
allow for stronger detections of this correlation.
According to our analysis, some experiments well suited to observing high CMB-multipoles like the KECK array, SPT, and SPIDER are not good candidates for observing FR. These instruments, while integrating to low noise depths, are observing in regions specifically selected for their low foregrounds and as seen in \autoref{fig: Radio RM}, FR is largest on the Galactic plane and falls off quickly as Galactic latitude increases.

Compared to single-frequency FR power spectrum estimators, like \citet{De:2013}, our expected signal-to-noise ratio using Planck-LFI data is greater by a factor of $\sim30$. The single-frequency estimators must also consider weak lensing effects of the CMB in order to accurately constrain their $\alpha^{RM}$ estimator. Since lensing contains only spatially dependent contributions \citep{Lewis:2006fu} and no dependence on frequency, the multi-frequency estimator considered here offers further advantages for reducing uncertainty from lensing. 
The use of single frequency or multifrequency estimators \citep{De:2013,pogosian:2013} is also able to estimate $\alpha^{RM}_{\ell m}$ from their techniques.
Although our cross-correlation results in the convolution of the polarized CMB and $\alpha^{RM}_{\ell m}$ power spectra, this work can be extended to provide an estimate of the FR power spectrum, $\alpha^{RM}_{\ell m}$, and to remove the effects of weak lensing on the FR estimator.

A map space analysis of non-Gaussain fields like FR will provide knowledge not accessible through the power spectrum. We can estimate the minimum polarized sensitivity required by a CMB experiment to observe detectable Galactic FR in a single pixel at 90 and 150~GHz using the maximum RM recovered by \citet{Oppermann:2014} of 2500 rad m$^{-2}$.

 We find that $\sim 1$ degree precision for polarization angle measurements is necessary to construct accurate maps of Galactic FR measure from CMB experiments. 
Even more sensitive observations are required to estimate extragalactic and primordial FR using CMB as a back-light. 
These sensitivities are currently beyond CMB experiments. 
\citet{planck:2015dust} discuss the uncertainty in
polarization angle which results from what they describe as conventional fitting and 
Bayesian inferencing. While some recovered angles show uncertainties
below this threshold, the distribution of uncertainties extends
up to 45 degrees for Bayesian-derived angles and 100 degrees for 
conventional estimation methods. These results are shown in Fig. B.1 in \cite{planck:2015dust}.

To achieve this level of sensitivity in a CMB experiment, we can 
estimate the uncertainty level an experiment would require on 
polarized emission by propagating the uncertainty in {\bf $\theta=\frac{1}{2}\tan^{-1}(\frac{U}{Q})$ }and assuming equality in uncertainties of $Q$ and $U$ ($\Delta Q = \Delta U$):
{\small \begin{equation}
\begin{aligned}
	\delta \theta^{2}& =(\frac{d\theta}{dQ} \Delta Q)^{2} + (\frac{d\theta}{dU}\Delta U)^{2} + 2 \frac{\delta\theta}{\delta Q} \frac{\delta \theta}{\delta U}\sigma_{QU}\\
	& = \frac{1}{4}\left(\frac{QU}{Q^{2}+U^{2}} \right)^{2} 
	\left( \left( \frac{\Delta Q}{Q}\right)^{2}  + \left(\frac{\Delta U}{U} \right)^{2}  \right) - \frac{1}{2} \frac{QU}{(Q^{2} + U^{2})^{2} } \sigma_{QU}\\
	&= \frac{1}{4}\frac{\Delta Q^{2}}{Q^{2} + U^{2}}  - \frac{1}{2} \frac{QU}{(Q^{2} + U^{2})^{2} } \sigma_{QU}\\
	\delta \theta^{2} &= \frac{1}{4} \left( \frac{\Delta Q^{2} }{P^{2}} - \frac{2QU\sigma_{QU}}{P^{4}} \right)
\end{aligned}
\end{equation}
}%

where $P =\sqrt{U^{2} + Q^{2}}$ is the polarization amplitude, and $\sigma_{QU}$ is the covariance between $Q$ and $U$.
This requires sensitivity at the $1--2\%$ level in polarization 
signal on a per pixel level when $\sigma_{QU} \to 0$.
In general, the presence of covariance between the observed $Q$ and $U$ Stokes parameters will complicate the ability to achieve sub-degree precision in polarization angle.

Through modelling and simulation, \citet{planck:2015pmf} predict that primordial magnetic fields of order 10nG will produce FR at comparable levels to galactic FR. For fields of these strengths, sub-degree polarization angle sensitivity and precise knowledge of galactic FR would be necessary to identify and characterize effects from these primordial fields.

Until these sensitivity levels are reached, a cross-correlation
can be used to identify the Galactic FR contributions to polarized power in the CMB. 
This kind of cross-correlation can also be used to verify the presence of FR from sources common to polarized surveys. 
Only the contributions to FR from common sources observed by both surveys [e.g. The CMB and \citet{Oppermann:2014} maps here] will produce a signal with this kind of correlation.

 In conclusion, we expect current
CMB experiments to be unable to detect FR even through cross-correlation. 
Strong residual signal from CMB foregrounds like dust and synchrotron radiation will need to be carefully removed from CMB in any analysis.
 AdvancedAct observations may be able to produce a statistically significant signal ($\sim 2$ sigma) through a cross-correlation and the significance of the signal will increase with the inclusion of higher multipole moments in the fitting of $\bar{\beta}$. 
Intermediary experiments that will have thermal noise similar to or lower than AdvancedAct may exhibit increasingly significant signals through this correlation. 
Strong detections will be possible with the construction of a future CMB-S4-type experiment.

}

\section{Acknowldegements}{
This research has been supported by Arizona State University.
We would like to thank Soma De, Tanmay Vachaspati, Daniel C. Jacobs,  Adam Beardsley and Samuel Gordon for their insightful conversations.
 Some of the results in this paper have been derived using the {\footnotesize HEALPIX} \citep{gorski_et_al2005} package.
 This research made use of {\footnotesize ASTROPY}, a community-developed core Python package for Astronomy \citep{astropy_2013}.

}

\bibliographystyle{mnras}
\bibliography{biblio}

\end{document}